\documentclass[showpacs,floatfix,aps]{revtex4}
\usepackage{amssymb,amsmath}
\usepackage{graphicx}
\usepackage{subfig}
\usepackage{caption}
\usepackage{float}
\usepackage[T1]{fontenc}
\usepackage{lscape}
\usepackage{cleveref}
\usepackage{lmodern}

\usepackage{lscape}
\def\be{\begin{equation}}
\def\ee{\end{equation}}
\def\ba{\begin{eqnarray}}
\def\ea{\end{eqnarray}}
\newcommand{\noi}{\noindent}

\begin{document}

\title{Acceleration of charged particles from near-extremal rotating black holes embedded in magnetic fields}

\author{C. H. Coimbra-Ara\'ujo}
\email{carlos.coimbra@ufpr.br}
\author{R. C. Anjos}
\email{ritacassia@ufpr.br}


\affiliation{Departamento de Engenharias e Ci\^encias Exatas,
  Universidade Federal do Paran\'a (UFPR),\\ 
Pioneiro, 2153, 85950-000 Palotina, PR, Brazil}
\affiliation{Applied physics graduation program, Federal University of Latin-American Integration, 85867-670, Foz do Iguassu, PR, Brazil.}



\begin{abstract}

The aim of the present article is to evaluate the motion of neutral and charged test particles in the vicinity of a near-extremal rotating black hole in the presence of magnetic fields. Euler-Lagrange motion equations and  effective potential methods are used to characterize the motion out of the equatorial plane. Such approach is of peculiar significance if it is considered, e.g., accretion processes onto rotating black holes. In general, investigations concerning accretion focus mostly on the simplest case of particles moving in the equatorial plane. Here it will be considered that particles initially moving around some particular orbit may be perturbed by a kick along the $\theta$ direction, giving rise to other possible orbits. We confirm the possibility that ultra high energy cosmic rays would be produced at the very center of AGNs, for a specific range of magnetic field magnitudes, since it is possible that ultra-high center-of-mass energies can be produced by particles colliding near the horizon of rotating black holes whose angular momentum tends to the Thorne limit. 

\end{abstract}

\maketitle


\section{Introduction}

Cosmic rays propagate in the universe in different energy ranges. Those with higher energy, frequently called {\it ultra high energy cosmic rays} (UHECRs), are detected in some terrestrial experiments like  Pierre Auger \cite{pierre} and Telescope Array \cite{ta} (TA) observatories. One of the main interests related to UHECRs is that their energy level is beyond the so called Greisen-Zatsepin-Kuzmin (GZK) limit ($E > 10^{18}$ eV). Accordingly, some important questions arise and one of the most intricated is related to the origin of these cosmic rays. The main explanations, depending on energy levels, are associated to mechanisms similar to that proposed by Fermi where charged particles could be accelerated by clouds of magnetized gas moving within our galaxy (aka Fermi's Mechanism). However Fermi's original mechanism is slow and inefficient to generate UHECRs and nowadays it was adapted to a diffusive shock acceleration mechanism  at shock fronts \cite {luis, federico}. A plethora of other proposes is also present in the literature (for a review see, e.g., \cite{vietri,olinto1,selvon,nagano,cronin}).

In any case, it is estimated that the acceleration of charged particles in the vicinity of the event horizon of central black holes (BHs) of active galaxy nuclei (AGNs) is one of the possible viable UHECR explanations. Astrophysical BHs have been observed by quite a few astronomical measurements as optical and X-ray observations and also gravitational wave detection. In this respect, there is a current endeavor among physicists and astronomers to answer the question if the BH theory studied by actual gravity theories (e.g., general relativity (GR) and modified theories of gravity) and astrophysical BHs are indeed the same thing. From this perspective, astrophysical BHs have some of the properties predicted by GR or other alternative theories concerning rotating BHs. Actually, most astrophysical objects have a mass and a spin and at least it appears that astrophysical BHs also have mass and spin, with angular momentum $a=J/M$ ($0<a<1$) \cite{parker,batal}. In particular, observations show that several stellar or supermassive BHs have high spin ($a\gtrsim 0.7$) \cite{bambi}. The outer region of these objects is described by some possible solutions of GR or other alternative theory. The classical example is a Kerr-type solution \cite{bardeen,kormendy,visser,teukolsky}. Besides, it is also important to consider magnetic fields since the traditional environment of AGNs consists of a plasma accretion disk surrounding the central black hole. Associated with such accretion phenomena there is the creation of highly relativistic jets and, as well, the dynamics of the accretion disk forms an electrodynamical system which produces magnetic fields as explained by several references \cite{blandford,garofalo,coimbraanjos,anjoscoimbra1,anjoscoimbra2}. Accordingly, the dynamics of the charged particle will be extremely sensitive to the presence of these fields, as will be discussed in Sections \ref{section:3} and \ref{section:4}.

Theoretically it is conceivable that UHECRs could be produced in the center of AGNs, since it is possible that ultra-high center-of-mass energies ($E_{c.m.}$) can be produced by particles colliding near the horizon of extremal rotating black holes ($a=1$). The literature presents the analyzes of such a possible situation in the vicinity of static black holes \cite{banados}, rotating black holes \cite{banados,jacobson,olvera}, charged black holes \cite{wei}, weakly magnetized black holes \cite{aliev,frolov,igata} and also Kerr naked singularities \cite{patil,stuchlik} (for a more complete list of references about the subject see \cite{pos2019}). Ultra high $E_{c.m.}$ particles were firstly proposed by Ba\~nados, Silk and West \cite{banados} who noticed that collisions of two neutral classical particles falling freely into extremal Kerr BHs ($a = 1$) may have infinite values of $E_{c.m.}$ close to the event horizon, if one of the particles is tracking marginally bound geodesics. In this respect, \cite{jacobson} and \cite{berti} concluded that, in fact, frame dragging effects in Kerr BHs can accelerate particle to high energies, but astrophysical restrictions on the spin (i.e., apparently, real black holes never reach $a = 1$) and restrictions on maximum $E_{c.m.}$ caused by gravitational radiation and back-reaction would solely permit infinite $E_{c.m.}$ only at infinite time and on the horizon of the black hole. About astrophysical limits on BH spin, supermassive BHs in the center of AGNs would have prolonged disk accretion and mass evolution thanks to galaxy merger events that, in both cases, yield black holes with very high spin \cite{bambi}. Nevertheless, the BH spin would be limited to $a \approx 0.998$ because the black hole would preferentially swallow negative angular momentum photons emitted by the accretion flow. This limit is also known as ``the Thorne limit'' (see, e.g., \cite{thorne,gammie,volonteri,benson,bambi2}). From now on, BHs with spin in the Thorne limit regime will be called ``near-extremal black holes'' \cite{brenneman,shi1,gralla,shi2}.  Although the statements pointed out by \cite{berti,jacobson} clearly indicates that the effects of self-force might prevent $E_{c.m.}$ from being arbitrarily high, it would also be feasible to undertake that the center of mass energy is still high enough to be of astrophysical interest if the mass-ratio of the point particle to the BH is negligible \cite{kimura,harada,igata}. This is particularly important for understanding UHECR production and acceleration process within AGN sources. 

In the present article, we study the collision of two particles (neutral-neutral, charged-neutral and charged-charged) with equal masses moving in the equatorial plane and also in other possible planes (where $0<\theta<90^\mathrm{o}$). The accelerated particles in the central environment of AGNs could mimic protons, atomic nuclei and other charged particles that could be considered as UHECRs. At the moment this work addresses particles as classical charged point particles, coming from infinity towards the horizon of Kerr black holes. Here the black hole is embedded in a magnetic field. Novel aspects are presented as the calculation of motion equations from a Lagrangian containing the electrodynamics characteristics that results in the interaction between the charged particle and the magnetic field, whose analysis is done in the equator position (as usually made) but also for different positions other than the equator. In other words, Euler-Lagrange motion equations and  effective potential methods are used to characterize the motion out of the equatorial plane. Here it will be considered that particles initially moving around some particular orbit may be perturbed by a kick along the $\theta$ direction, giving rise to other possible orbits. This means that, in the context of polar coordinates, $\theta$ varies with $r$ and the geodesics $\dot{\theta}$ is non null. In this way, the geodesics $\dot{t}$, $\dot{r}$, $\dot{\theta}$ and $\dot{\phi}$ are computed, resulting in the calculation of the effective potential $V_{eff}$ and the center of mass energy $E_{c.m.}$ of two particles that collide close to the horizon. It is possible to assess how much the system is bond or not and if there is a possibility of particles escape quickly from the system. It is concluded from the results that in fact rotating black holes function as a kind of slingshot propelling the particles out of the system if: a) there are no magnetic fields and the black hole is in the extremal regime ($a=1$, as showed by \cite{banados}); b) the magnetic field is on and the black hole is in the near-extremal regime ($a \approx 0.998$). In this last case, only if one of the particles is charged, magnetic fields work by mimicking the effect of black hole spin $a$. It is important to point that only a certain range of magnetic fields actually cause the acceleration of the particle in positions increasingly farther from the horizon (Section \ref{section:4}). Otherwise, magnetic fields of great magnitude can cause the particle to be trapped in the system, prompting the particle to fall into the black hole. In addition, the axial acceleration of the particle is evaluated, showing that for certain fields and spins, in fact the charged particle tends to accelerate out of the system. Analyzes in future works will include the fact that these particles are also fermions, excluding cases applied to small black holes (where Hawking or Unruh effect would be applicable) since we are dealing with astrophysical black holes, of great mass and large area of the horizon. 

This article is divided as follows. In Section \ref{section:2} we explain particle energy from Kerr background. Section \ref{section:3} is dedicated to discuss electrically charged particles in the presence of magnetic fields and the impact of this in particle geodesics. Section \ref{section:4} presents the main results for geodesics and for the energy of neutral-neutral, neutral-charged or charged-charged particle collisions near Kerr BHs. In this section some discussion is presented as well. In Section \ref{section:concluding} we present some concluding remarks. Here it will be considered $G=c=1$ and metric signature $(-+++)$.

\section{Spacetime background and particle energy}\label{section:2}

The Kerr BH is described by two parameters: its mass $M$ and its angular momentum $J$ (here represented by $a=J/M$, i.e., the angular momentum per unit of mass). The Kerr line element describes a stationary spacetime with axial symmetry and, in Boyer-Lindquist coordinates, it is written as \cite{teukolsky}

\be\label{eq:kerr}
ds^2 = g_{tt}dt^2 + 2g_{t \phi}dtd\phi + g_{rr}dr^2 + g_{\theta \theta}d\theta^2 + g_{\phi \phi}d\phi^2,
\ee

\noi with

\be
g_{t t} = -\left(1-\frac{2M}{\Sigma} \right),
\ee

\be
g_{t \phi} = -\frac{2aMr\sin^2\theta}{\Sigma},
\ee

\be
g_{r r} = \frac{\Sigma}{\Delta},
\ee

\be
g_{\theta \theta} = \Sigma,
\ee

\be
g_{\phi \phi} = \frac{(r^2+a^2)^2-a^2\Delta \sin^2\theta}{\Sigma}\sin^2\theta,
\ee

\noi and where $\Sigma=r^2+a^2\cos^2\theta$ and $\Delta = r^2+a^2-2Mr$. The event horizon is located at $r_H = M + \sqrt{M^2 - a^2}$. Since static holes have no rotation, then $a\rightarrow 0$ and the Kerr horizon coincides with the Schwarzschild one $r_H = r_S = 2M$. The so called ergoregion is described by $r_H < r < r_E(\theta)=M+\sqrt{M^2-a^2\cos^2\theta}$. From here, we will consider $M=1$. Note that in this situation the horizons are $r_H = 2$ for $a=0$ (static BH) and $r_H = 1$ for $a=1$ (extremal BH). \\

\begin{figure}[h]
  \centering
  \includegraphics[width=8cm]{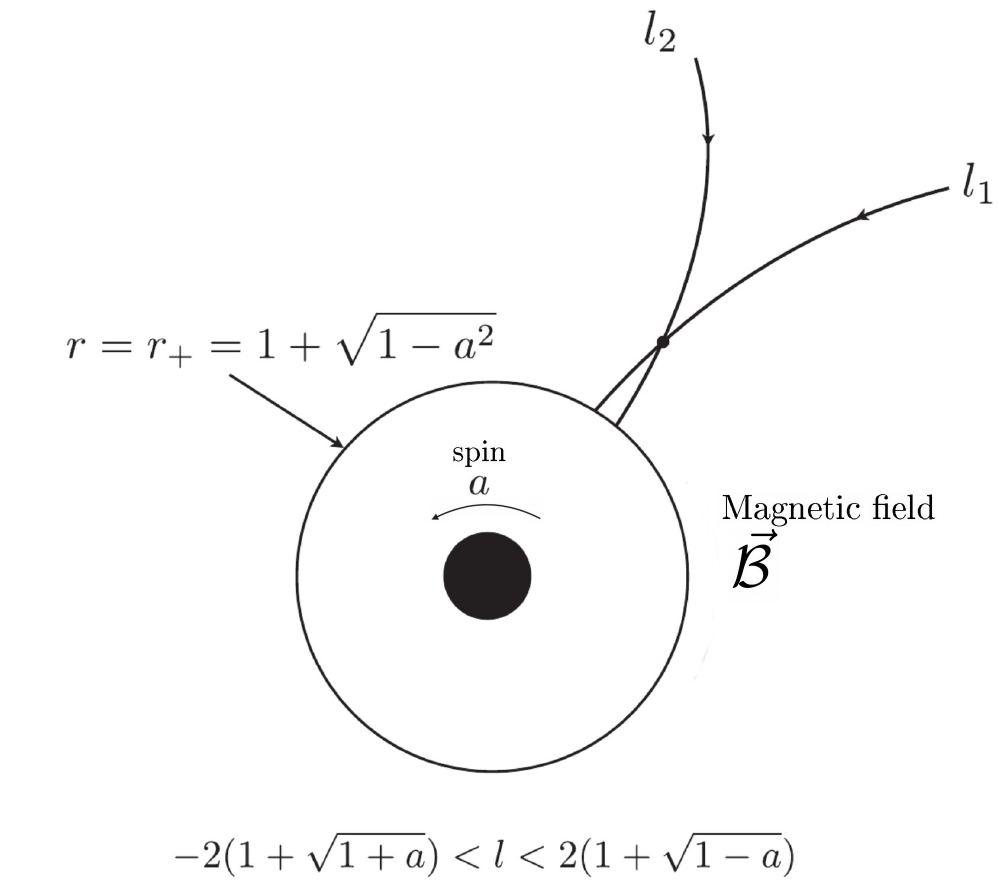}\\
\caption{Rotating black hole with magnetic fields near its horizon. Here when a charged particle is orbiting in the innermost stable circular orbit (ISCO), the last stable circular orbit around  BHs, of a rotating black hole, an incoming neutral or charged particle can shot it, expelling it to infinity. Figure adapted from \cite{banados}.}
\label{fig:fig1}
\end{figure}

When considering the motion of particles only under the action of gravity, the conserved quantities along geodesics play a fundamental role. The energy of the particle, as well as
its angular momentum relative to the axis of symmetry, are conserved quantities as a consequence of Kerr metric symmetries and the Noether theorem. Considering the motion of neutral or charged particles near rotating BHs in a background described by (\ref{eq:kerr}), i.e., a vacuum rotating black hole, the conserved quantities are attached to Killing vectors $\xi_{(t)}=\xi_{(t)}^\mu\partial_\mu=\frac{\partial}{\partial t}$ and $\xi_{(\phi)}=\xi_{(\phi)}^\mu\partial_\mu=\frac{\partial}{\partial \phi}$. The first one is related to the free test particle energy conservation 

\be\label{eq:killing1}
\mathcal{E}=-g_{t\mu}p^\mu, 
\ee

\noi and the other to the free test particle angular momentum conservation

\be\label{eq:killing2}
\ell=-g_{\phi\mu}p^{\mu}.
\ee

\noi Here the range of $\ell$, the angular momentum per unit rest mass, for geodesics falling in is $-2(1+\sqrt{1+a})<\ell <2(1+\sqrt{1-a})$ \cite{banados}. 

When two neutral or charged particles approach the horizon of a black hole, since the background is curved, it is necessary to define the center-of-mass frame properly. The center-of-mass energy of the two particle system (each with mass $m_0$) is given by \cite{banados}

\be\label{eq:ecm}
E_{c.m.} = m_0 \sqrt{2} \sqrt{1-g_{\mu \nu}u^\mu_{(1)}u^\nu_{(2)}},
\ee

\noi where $u^\mu_{(1)}$ and $u^\nu_{(2)}$ are the 4-velocities of each particle, properly normalized by $g_{\mu\nu}u^\mu u^\nu=-1$. It is awaited that the main conditions for accelerating the particles are regarded from the BH spin, since spin influences the position of the innermost stable circular orbit (ISCO), the last stable circular orbit around the BH. The idea is that the greater the spin $a$ is the more displaced the minimum of the potential energy becomes, since the stability region depends only on the rotation parameter of the BH \cite{aliev}. This displacement occurs in the direction of the horizon, indicating that the ISCO is closer to the horizon as the value of the spin $a$ approaches 1. We will show in next sections that magnetic fields also could affect the particle motion. A sketch of the system is shown in Fig. \ref{fig:fig1}.

\section{Electrically charged particles in the presence of magnetic fields}\label{section:3}

Many observations point that a magnetic field is present in the center of AGN environment.  Astrophysically, the origin of such a magnetic field is explained from plasma motion connected to accretion disks \cite{mckinney}. In particular, the observation of the twin-jet system of NGC 1052 at 86 GHz with the Global mm-VLBI Array, derived the magnitude of magnetic fields at the very center of AGNs: it would be between $\mathcal{B} \sim 200$ G and $8.3 \times 10^4$ G at 1 Schwarzschild radius \cite{baczko}. See also \cite{piotrovich,eatough,shannon} for similar estimations.

In this context, it is supposed that magnetic fields at the AGN core could affect and alter the motion of charged particles that are orbiting the central black hole \cite{znajek}. When a charged particle is orbiting in the ISCO of a rotating black hole, an incoming particle (here we are considering or a neutral or a charged one) can shot it, leading to three possible outcomes (analyzed from center of mass point of view): the charged particle is expelled to infinity, or it could be trapped by the BH, or it converges to stability and persists orbiting in the ISCO.

For a particle immersed in an uniform magnetic field in a curved space, its Lagrangian is

\be\label{lagrange}
\mathcal{L} = \frac{1}{2}g_{\mu \nu} \dot{x}^{\mu} \dot{x}^\nu + \frac{qA_{\mu}\dot{x}^{\mu}}{m_0}, 
\ee

\noi where $A_\mu$ is the electromagnetic 4-potential, $q$ is the particle charge and $m_0$ is the particle mass. The 4-potential is invariant under the symmetries which correspond to the Killing vectors, i.e., $A_{\mu,\nu} \xi^\nu + A_\nu\xi^\nu_{,\mu}=0$.  The fields $\xi^\nu$ and $\xi^\nu_{,\mu}$ are  used  for describing electromagnetic fields in the Kerr metric. They are superpositions of Coulomb electric and uniform magnetic fields. Respecting the Lorentz gauge $A^\mu_{;\mu}=0$ one can chosen for example \cite{aliev}

\be
A_\mu = \left(\frac{2am_0 B}{q},0,0,\frac{m_0 B}{q} \right),
\ee 

\noi where $B = \frac{q\mathcal{B}}{2m_0}$. The 4-momentum of the particle in this case is $p_\mu = m_0 u_\mu + q A_\mu$ and the conserved quantities in eqs. (\ref{eq:killing1}) and (\ref{eq:killing2}) are written as

\be\label{eq:newkilling1}
\mathcal{E}=-g_{t\mu}(m_0 u^\mu + q A^\mu), 
\ee

\be\label{eq:newkilling2}
\ell=-g_{\phi\mu}(m_0 u^\mu + q A^\mu).
\ee

\section{Main results and discussion}\label{section:4}

The Euler-Lagrange equations from Lagrangian (\ref{lagrange}), considering the conserved quantities in (\ref{eq:newkilling1}) and (\ref{eq:newkilling2}), where $\ell$ is the angular momentum of the particle and $\mathcal{E}=1$ the energy, and the normalization condition $u^\mu u_\mu = -1$, lead to the following system of equations where it is possible to calculate the solution for the geodesic equations $\dot{t}$, $\dot{\phi}$, $\dot{r}$ and $\dot{\theta}$:

\be
\dot{t}=-\frac{1}{r^2}[a(aE(r,\theta)-L(r,\theta)) + (r^2+a^2)T/\Delta],
\ee

\noi and

\be
\dot{\phi}=-\frac{1}{r^2}[(aE(r,\theta)-L(r,\theta)) + aT/\Delta],
\ee

\be\label{1}
-\frac{1}{2}\left(\frac{\partial g_{r r}}{\partial \theta}\right) \dot{r}^2 -\frac{1}{2}\left(2a^2\cos\theta \sin\theta + \frac{\partial g_{\theta \theta}}{\partial \theta}\right) \dot{\theta}^2 +2r\dot{r}\dot{\theta} = \frac{1}{2}\left(\frac{\partial g_{t t}}{\partial \theta}\right) \dot{t}^2 + \frac{1}{2}\left(\frac{\partial g_{\phi \phi}}{\partial \theta}\right) \dot{\phi}^2 + \left(\frac{\partial g_{t \phi}}{\partial \theta}\right) \dot{t}\dot{\phi},
\ee

\be\label{2}
g_{r r} \dot{r}^2 + g_{\theta \theta} \dot{\theta}^2 = -1 - g_{t t} \dot{t}^2 - g_{\phi \phi} \dot{\phi}^2 - 2g_{t \phi} \dot{t}\dot{\phi},
\ee

\noi where $E(r,\theta)= 1- 2aB +2aBr/\Sigma$, $L(r,\theta) = \ell + 2a^2 Br/\Sigma  - (r^2+a^2)B$, $T(r,\theta) = E(r,\theta) (r^2 + a^2) -L(r,\theta) a$.

\noi In particular, the system of equations (\ref{1}) e (\ref{2}) for geodesics $\dot{r}$ and $\dot{\theta}$ is solved as

\be
\dot{r} = \pm \sqrt{\frac{-W\pm\sqrt{W^2 - 4YF^2}}{2Y}},
\ee

\be
\dot{\theta} = \sqrt{G - \frac{g_{rr}}{g_{\theta \theta}}\dot{r}^2},
\ee

\noi where

\be
Y =  \frac{\partial g_{r r}}{\partial \theta} + 16 r^2 \frac{g_{rr}}{g_{\theta \theta}},
\ee

\be
W = 2\frac{\partial g_{r r}}{\partial \theta}F - 16r^2G,
\ee

\be
F = \left(\frac{\partial g_{t t}}{\partial \theta}\right) \dot{t}^2 + \left(\frac{\partial g_{\phi \phi}}{\partial \theta}\right) \dot{\phi}^2 + 2\left(\frac{\partial g_{t \phi}}{\partial \theta}\right) \dot{t}\dot{\phi},
\ee

\be
G = \frac{-1 - g_{t t} \dot{t}^2 - g_{\phi \phi} \dot{\phi}^2 - 2g_{t \phi} \dot{t}\dot{\phi}}{g_{\theta \theta}}.
\ee

\noi The dot refers to proper time derivatives. 

\begin{figure}%
    \centering
    \subfloat[]{{\includegraphics[width=8cm]{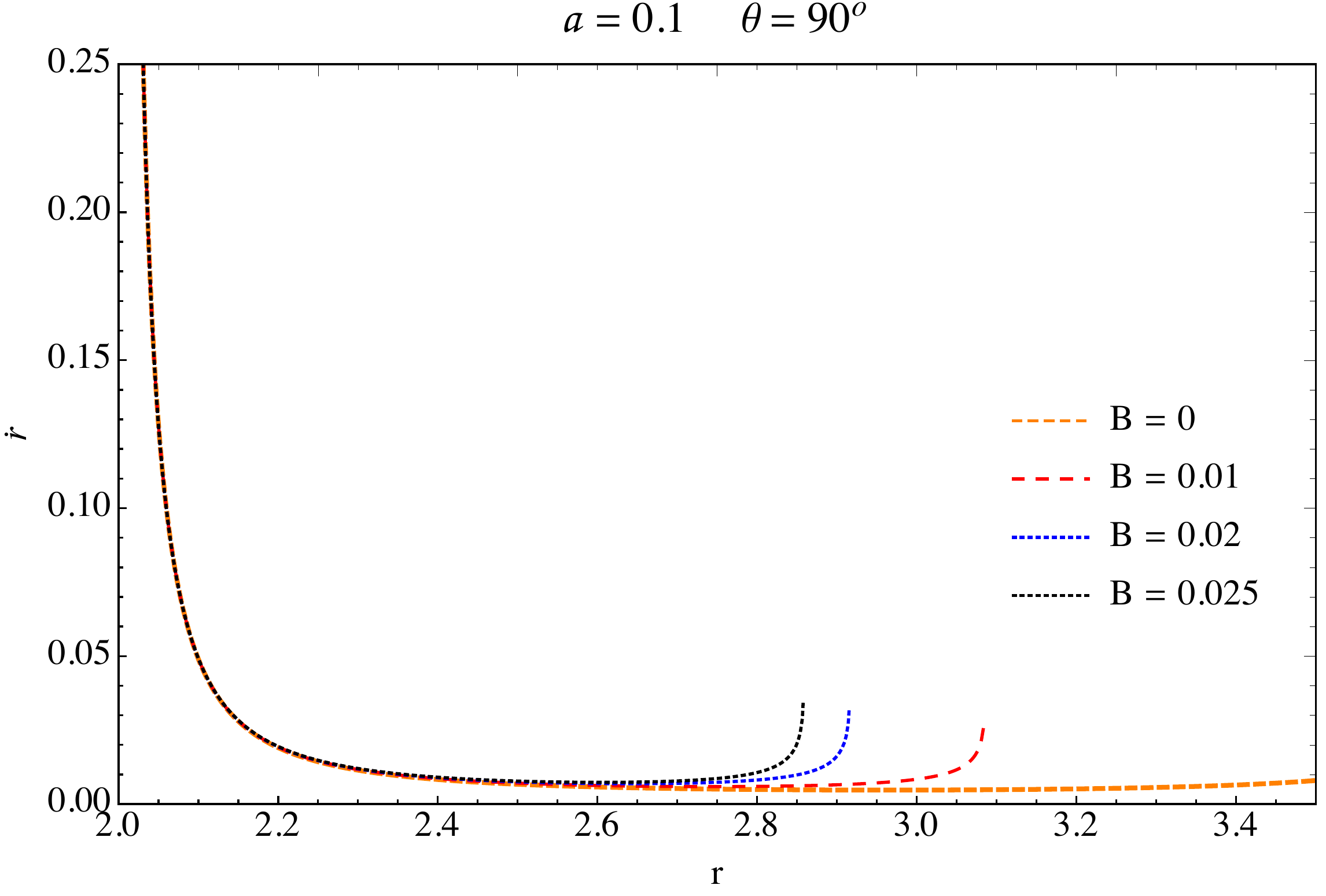} }}%
    \qquad
    \subfloat[]{{\includegraphics[width=8cm]{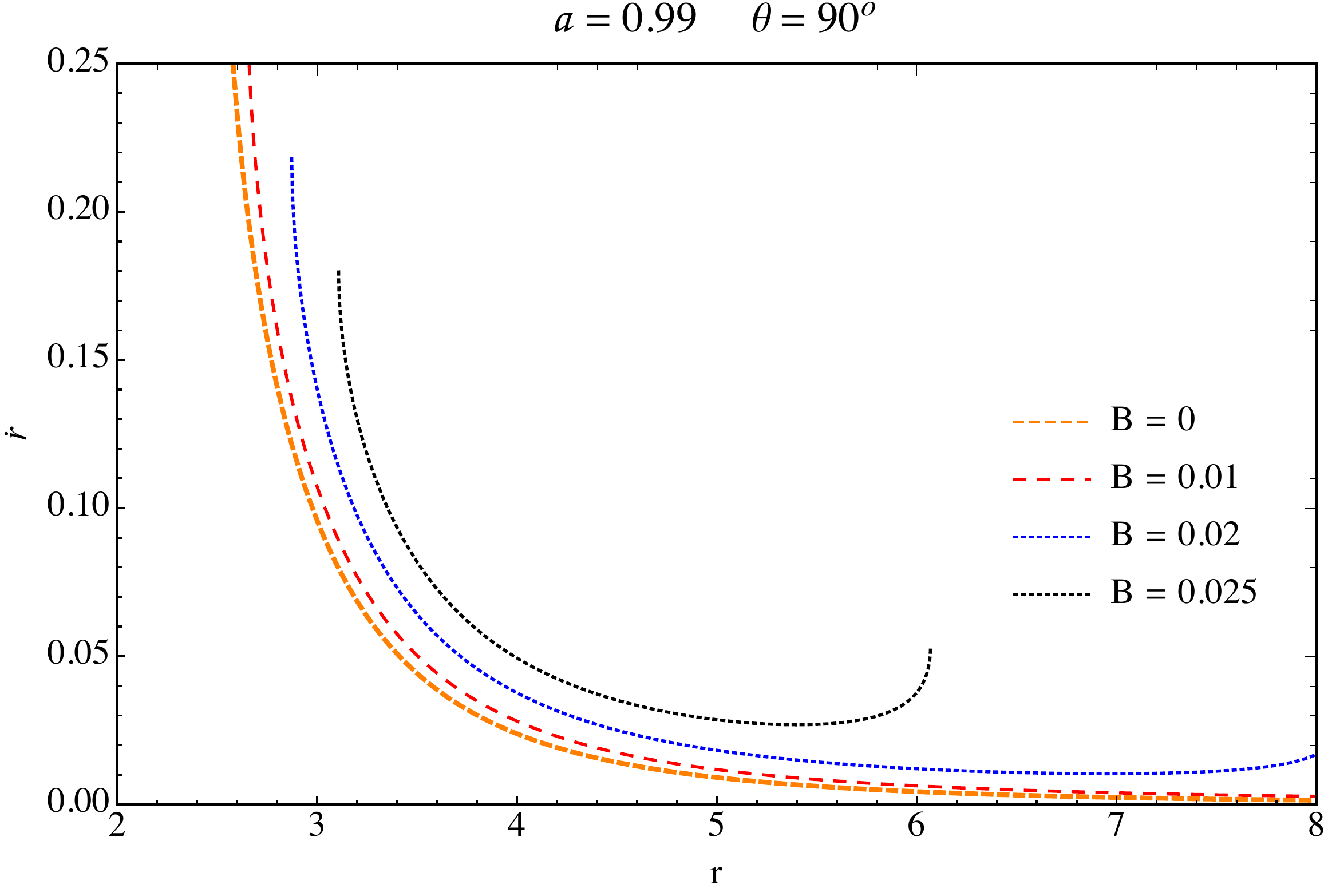} }}%
    \caption{(a) Radial velocity for $a=0.1$ and some values of $B$ in C.T/m units. (b) The same for $a=0.9$.}%
    \label{fig:fig2}%
\end{figure}

\begin{figure}%
    \centering
    \subfloat[]{{\includegraphics[width=8cm]{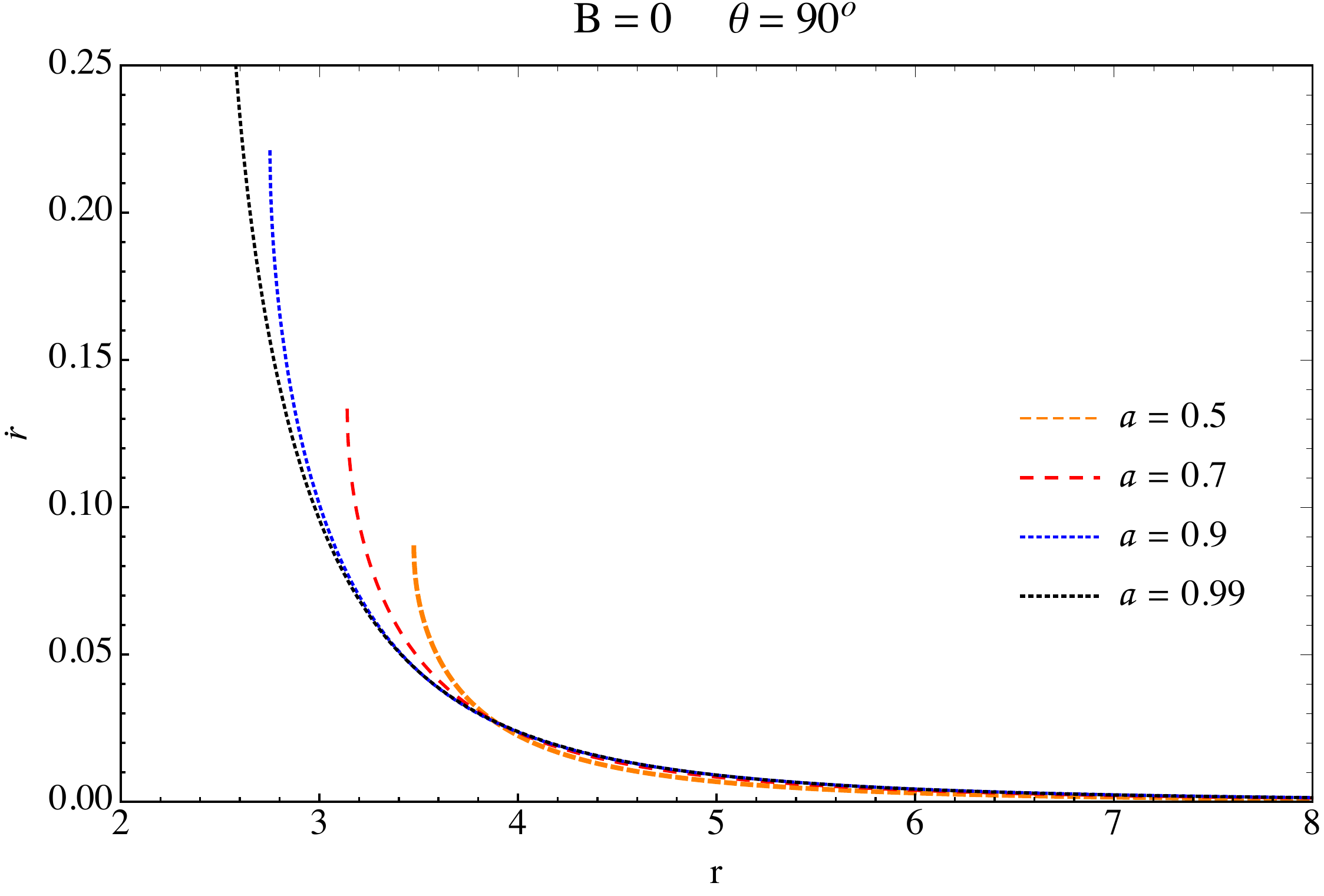} }}%
    \qquad
    \subfloat[]{{\includegraphics[width=8cm]{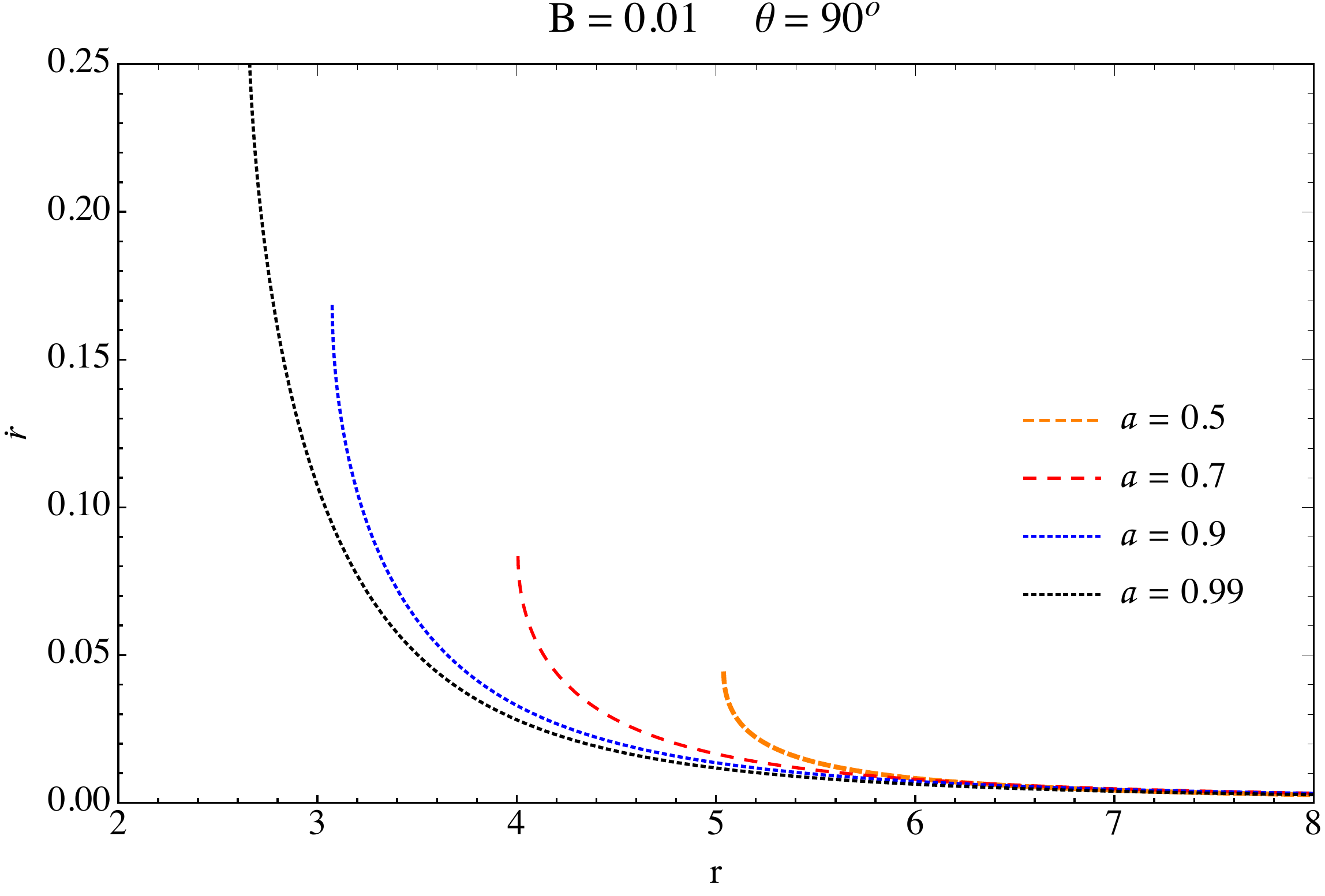} }}%
    \caption{(a) Radial velocity for $B=0$ and some values of $a$. (b) The same for $B=0.01$ in C.T/m units.}%
    \label{fig:fig3}%
\end{figure}

\begin{figure}%
    \centering
    \subfloat[]{{\includegraphics[width=8cm]{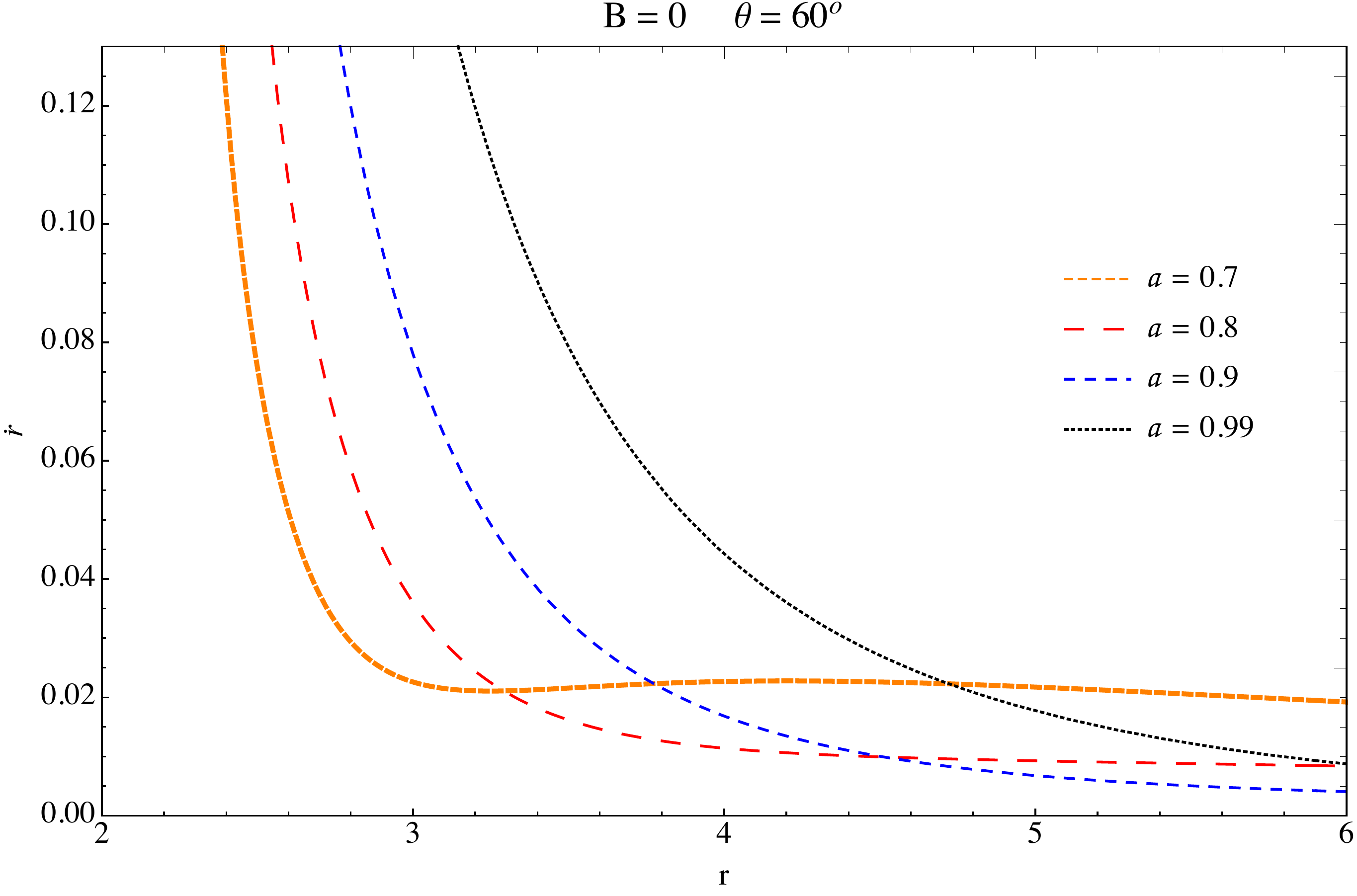} }}%
    \qquad
    \subfloat[]{{\includegraphics[width=8cm]{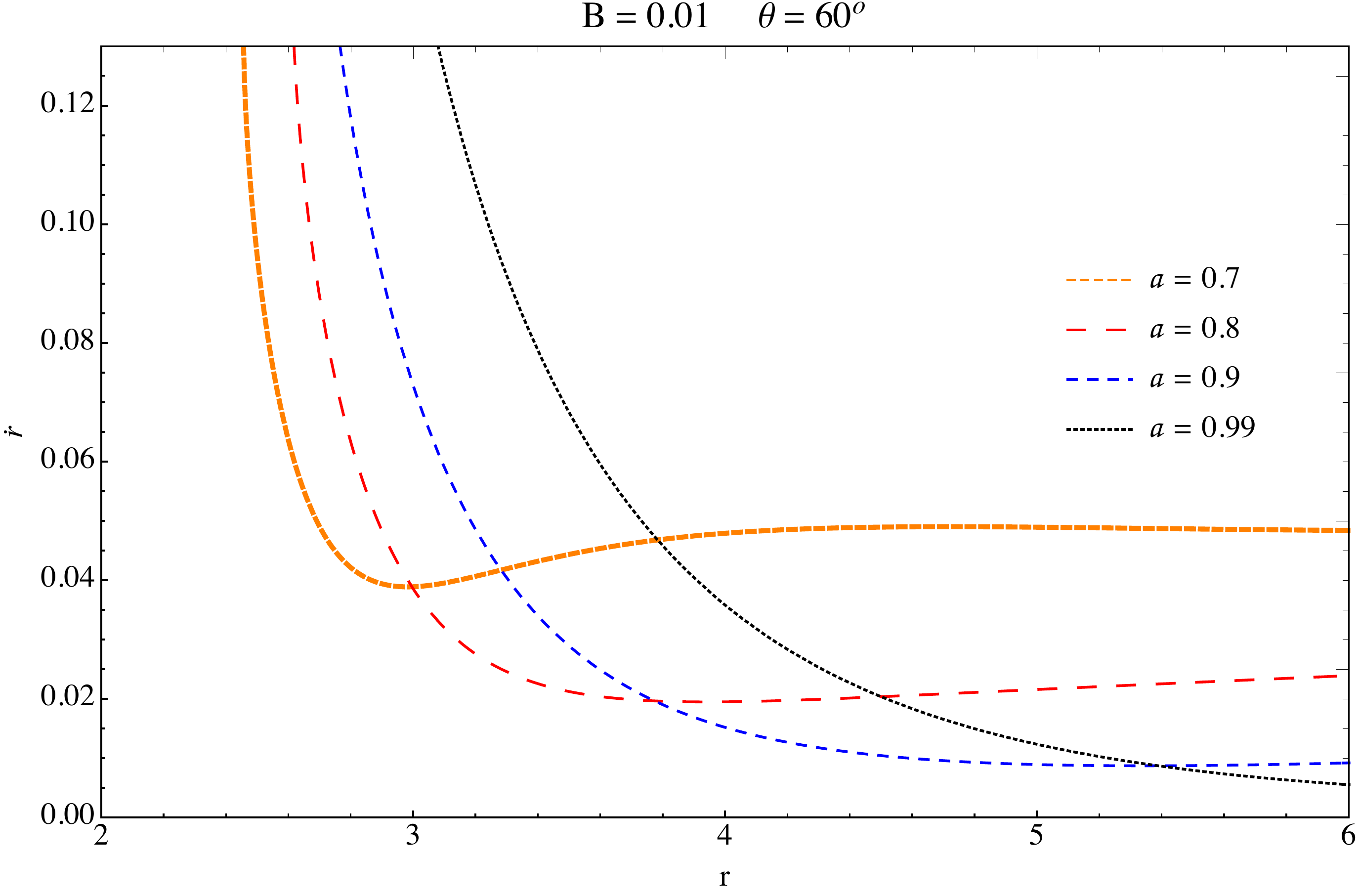} }}%
    \caption{(a) Radial velocity for $B=0$ out of the equatorial plane and for some values of $a$. (b) The same for $B=0.01$, in C.T/m units.}%
    \label{fig:fig4}%
\end{figure}

\begin{figure}%
    \centering
    \subfloat[]{{\includegraphics[width=8cm]{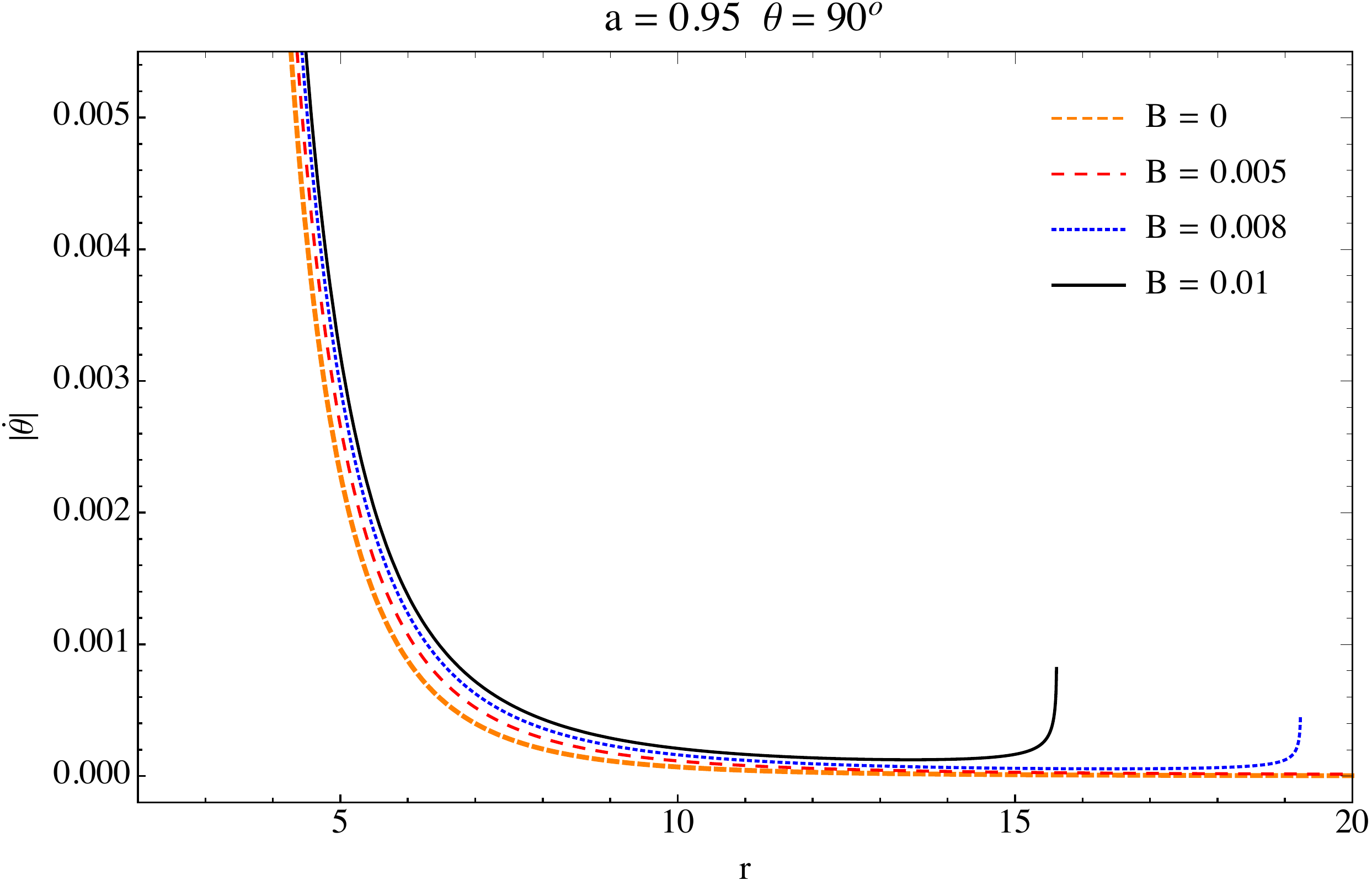} }}%
    \qquad
    \subfloat[]{{\includegraphics[width=8cm]{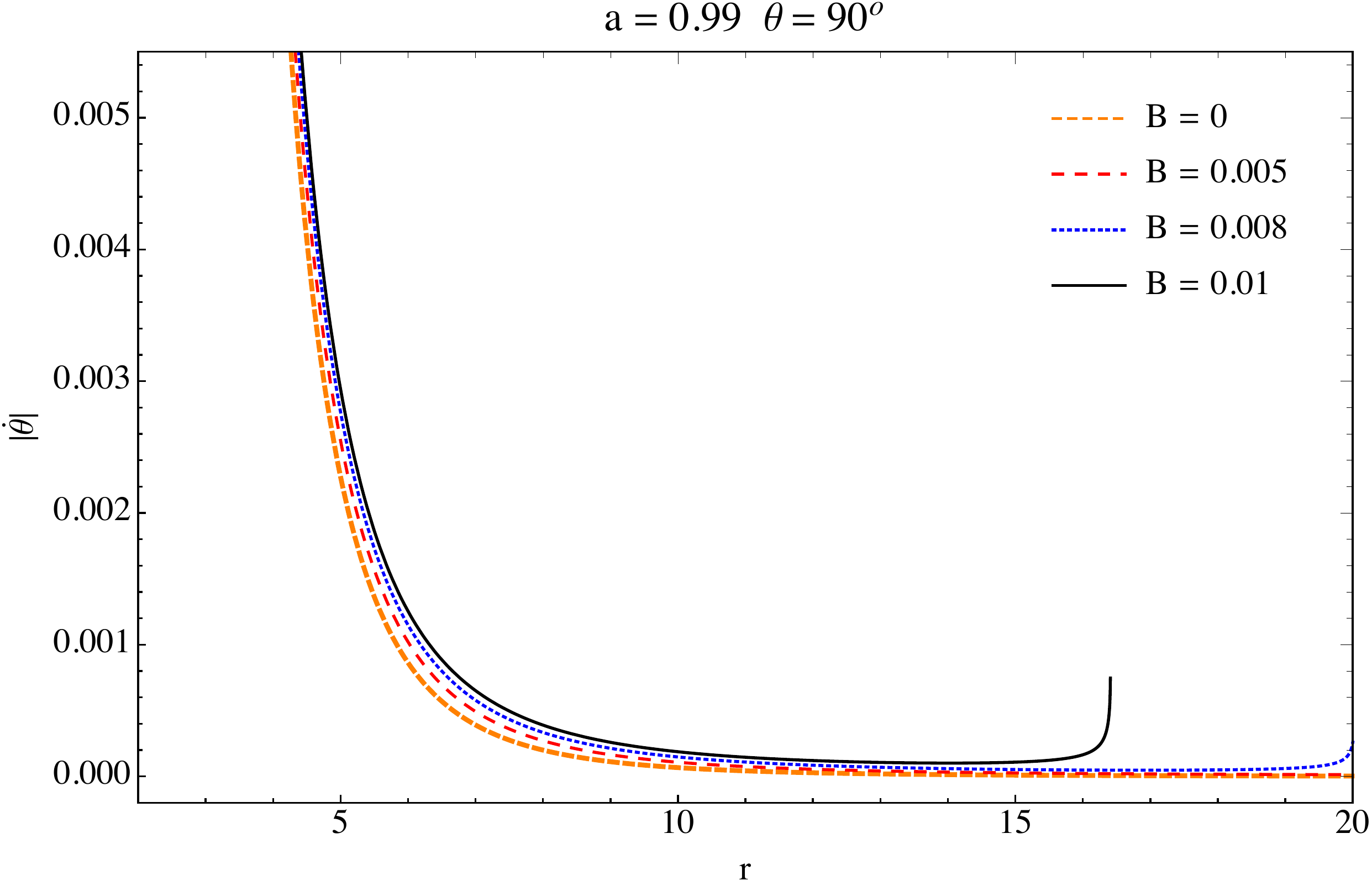} }}%
    \caption{(a) Axial angular velocity for $a=0.95$ and some values of $B$ in C.T/m units. (b) The same for $a=0.99$.}%
    \label{fig:fig5}%
\end{figure}

Figs. \ref{fig:fig2}, \ref{fig:fig3} and \ref{fig:fig4} show the variation of radial geodesic $\dot{r}$ for values of $a$ and $B$, in the equatorial plane or not. Regarding the values of $B=q\mathcal{B}/2m_0$ hereon we are considering $q/m_0 = 1$ C/m (in natural units). As seen at the beginning of Section \ref{section:3}, at the vicinity of the black hole the value of the magnetic field is $\mathcal{B} \sim 10^2-10^4$ G (i.e. $\sim 10^{-2}-1$ T). Thus, here we are using $\sim 10^{-2}-1$ C.T/m as the range of values for B.  The results show that the trajectory of particles is very sensitive to the presence of magnetic fields, even for small values of the magnetic fields. Fig. \ref{fig:fig5} shows the behavior of $\dot{\theta}$ geodesics for four different values of magnetic fields. The action of the magnetic field on motion in Figs. \ref{fig:fig2}-(b), \ref{fig:fig3}-(b) and \ref{fig:fig5} is to accelerate the particle, indicating the possibility that Kerr black holes can act as natural accelerators. On the other hand, the radial motion for particles moving along geodesics out the equatorial plane ($\theta = \pi/3$) show that particles are found to get trapped by the BHs if the value of magnetic fields and BH spin are small (see Fig. \ref{fig:fig4}). 

For a test particle with mass $m_0$, the effective potential $V_{eff}(r)$ is calculated from radial geodesics as

\be
\frac{1}{2}\dot{r}^2 + V_{eff}(r) = 0.
\ee

\noi The effective potential $V_{eff}(r)$ is then

\be
V_{eff}(r) = - \frac{1}{2}\dot{r}^2  = -\frac{1}{2}\Biggl(\sqrt{\frac{-W\pm\sqrt{W^2 - 4YF^2}}{2Y}}\Biggl)^ {2}.
\ee

\begin{figure}[h]
  \centering
  \includegraphics[width=8cm]{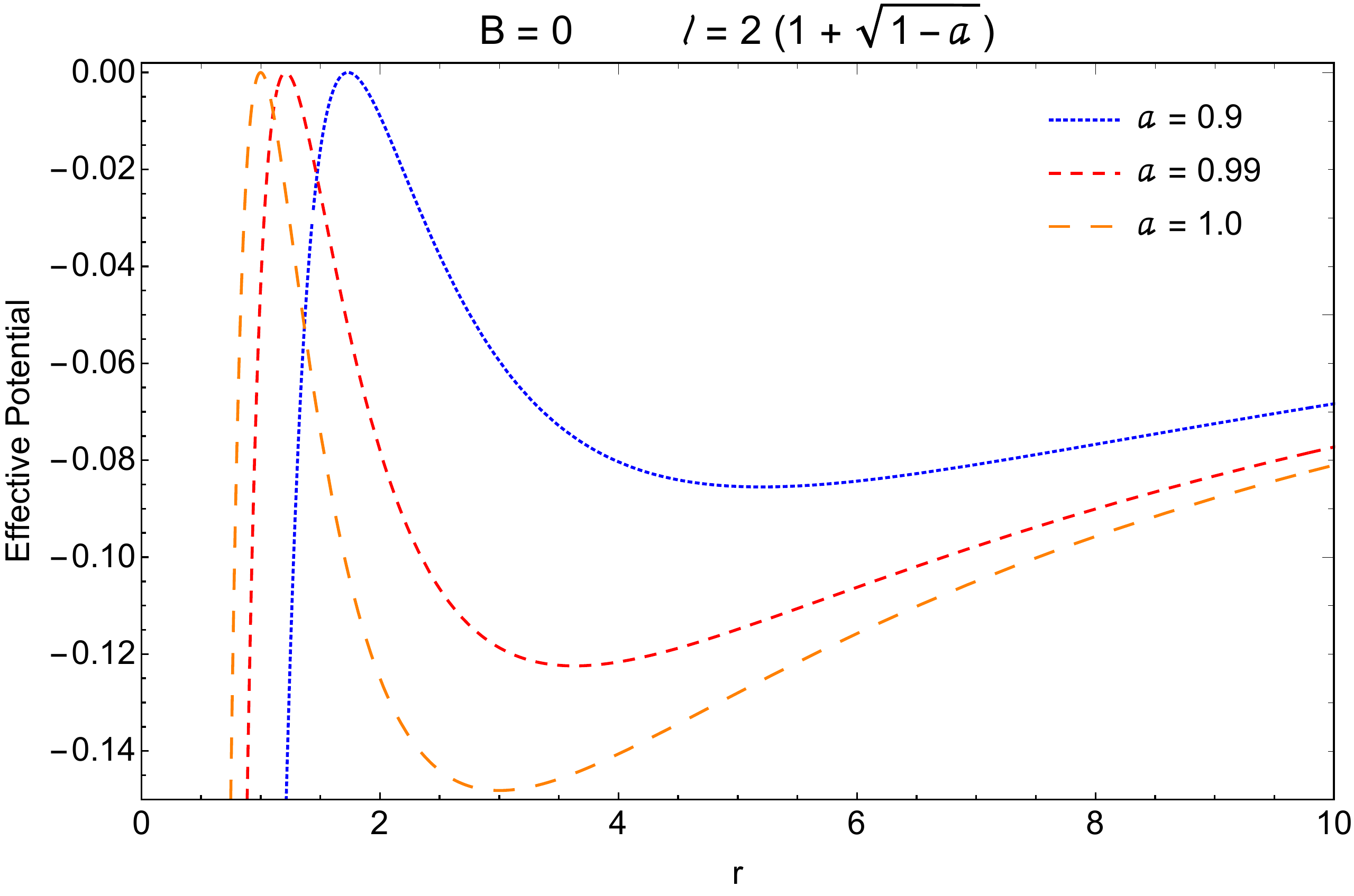}\\
\caption{Effective potential for marginally bound critical particles (no magnetic field situation).  The potential must be non-positive  in  the  allowed  region  of  particle  motion.  Here the test particle develops its maximum angular momentum $\ell=2(1+\sqrt{1-a})$. The  minimum in the potential  describes  where  the  ISCO  is located for some possible spin $a$. The greater the spin is the closer to the horizon the ISCO becomes.}
\label{fig:fig6}
\end{figure}

\begin{figure}%
    \centering
    \subfloat[]{{\includegraphics[width=8cm]{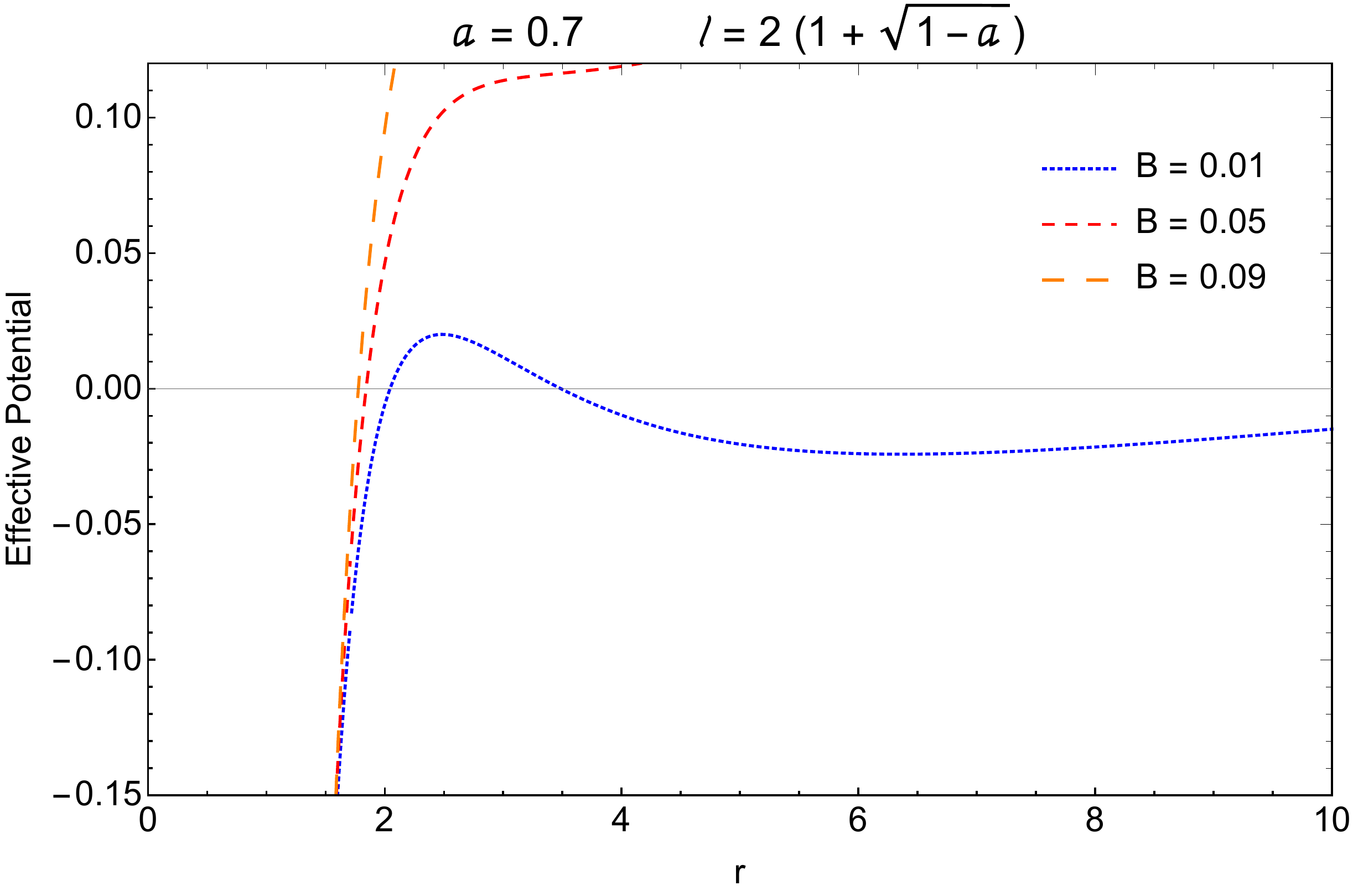} }}%
    \qquad
    \subfloat[]{{\includegraphics[width=8cm]{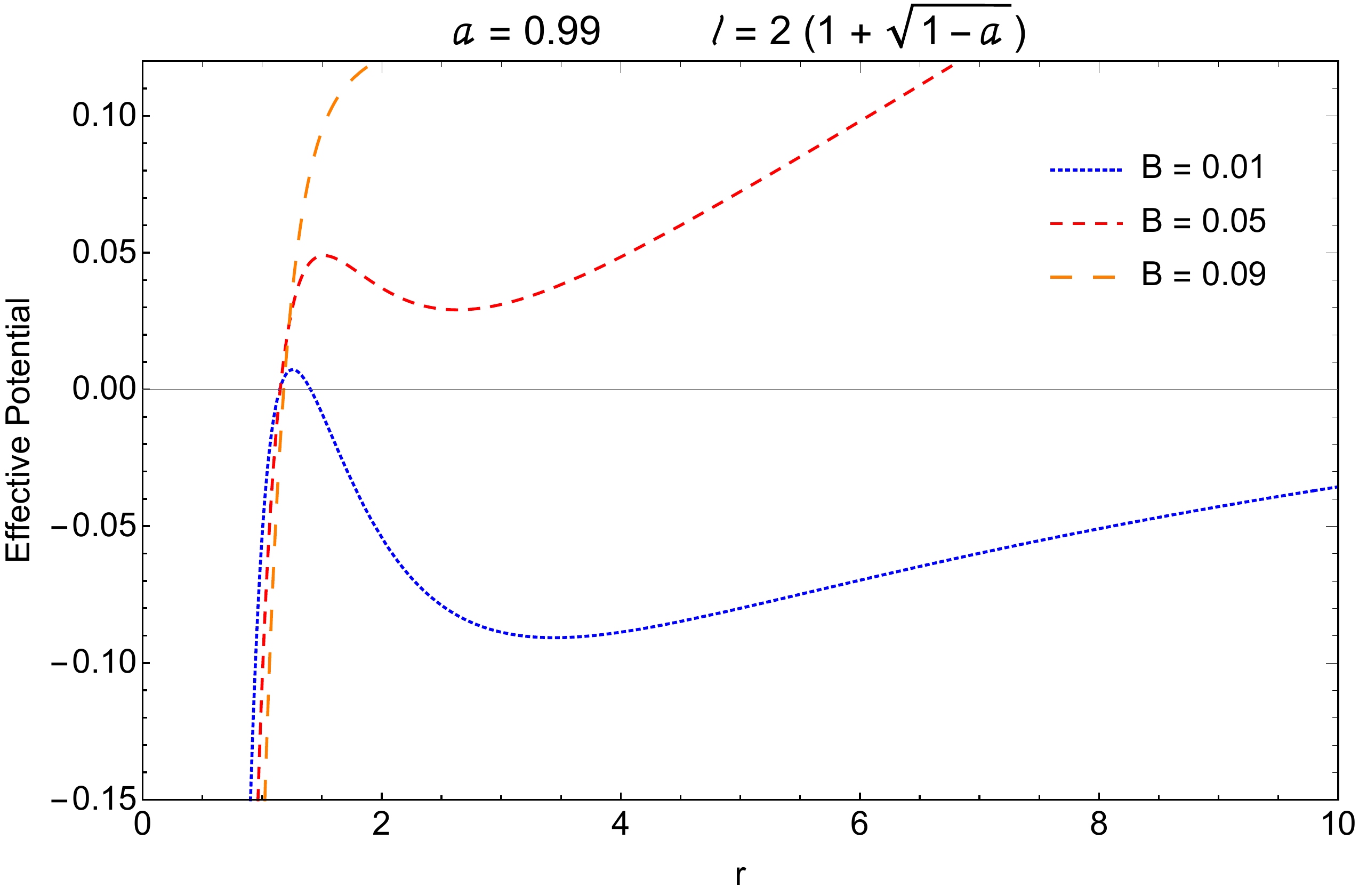} }}%
    \caption{The presence of magnetic fields changes the position of ISCO. The greater the magnetic field magnitude is the farther the ISCO position become. In this situation, the ISCO particle is accelerated from positions more distant of the horizon than if there is no magnetic field. It is confirmed from results plotted in Fig. \ref{fig:Ecm}. $B$ in C.T/m units.}%
    \label{fig:fig7}%
\end{figure}

\noi The  minimum of potential $V_{eff}(r)$ describes  where  the  ISCO  is located for some possible spin $a$. The effective potential is shown in Fig. \ref{fig:fig6}, with no magnetic field, and in Fig. \ref{fig:fig7} when magnetic field is set on. Note that magnetic field values above a certain limit destabilizes the orbit of the particles. Regarding the relation between $V_{eff}$ and the spin $a$, the greater the spin is the closer to the horizon the ISCO become. In this case, the $r_{isco}$ comes from

\be
\frac{\partial V_{eff}}{\partial r} = 0,
\ee

\noi and

\be
\frac{\partial^2 V_{eff}}{\partial r^2} \leq 0,
\ee

\begin{figure}[h]
  \centering
  \includegraphics[width=8cm]{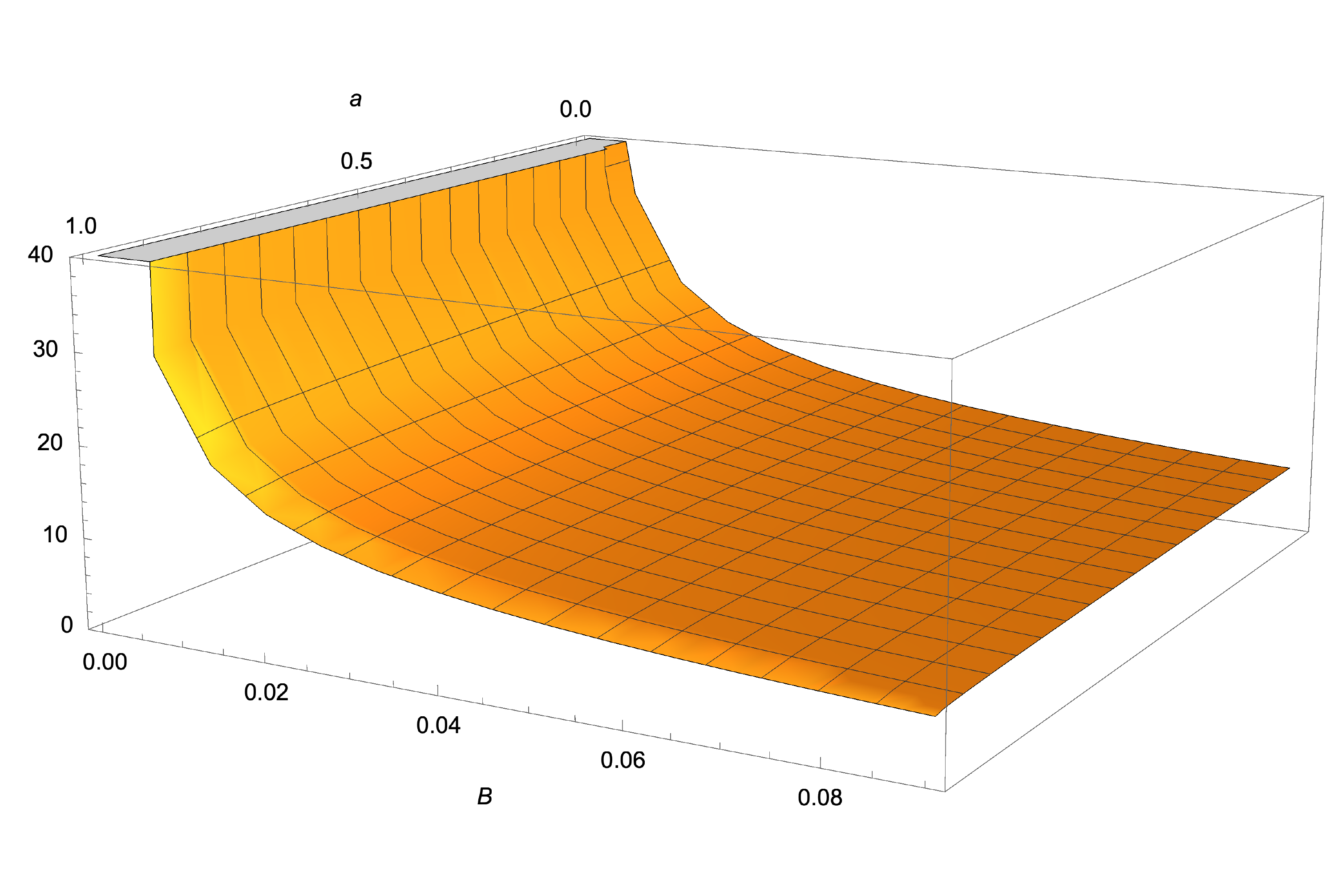}\\
\caption{Dependence of the radii of the stable circular orbits as function of spin $a$ and magnetic field $B$.}
\label{fig:fig8}
\end{figure}

\noi and its relation with the spin $a$ and the normalized magnetic field $B$ is shown in Fig. \ref{fig:fig8}. The radii os stable orbits changes considerably with small increases in $B$ (the greater $B$ is, the smaller the value of $r_{isco}$ becomes, exponential relation). The same inversely proportional relationship occurs between $a$ and $r_{isco}$, but a smoother relation. This allows us to say that magnetic fields roughly work mimicking the effect of black hole spin $a$.

Calculation of center of mass energy from particle collision comes from eq. (\ref{eq:ecm}), with

\begin{equation}
g_{\mu \nu}u^\mu_{(1)}u^\nu_{(2)} = \left(\frac{2}{r} - \frac{a^2}{r^2} -1 \right)\frac{T_1 T_2}{\Delta^2} + \frac{(aE_1 - L_1)(aE_2 - L_2)}{r^2} + \frac{r^2}{\Delta} \dot{r}_{(1)}\dot{r}_{(2)},
\end{equation}

\noi where

\be
E_1(r,\theta)= E_2(r,\theta) = 1- 2aB +2aBr/\Sigma,
\ee

\be
L_1 (r,\theta) = \ell_1 + 2a^2 Br/\Sigma  - (r^2+a^2)B, ~~~~~~~~~~ L_2 = \ell_2 + 2a^2 Br/\Sigma  - (r^2+a^2)B,
\ee

\be
T_i (r,\theta) = E_i (r,\theta) (r^2 + a^2) -L_i (r,\theta) a.
\ee

\begin{figure}%
    \centering
    \subfloat[]{{\includegraphics[width=8cm]{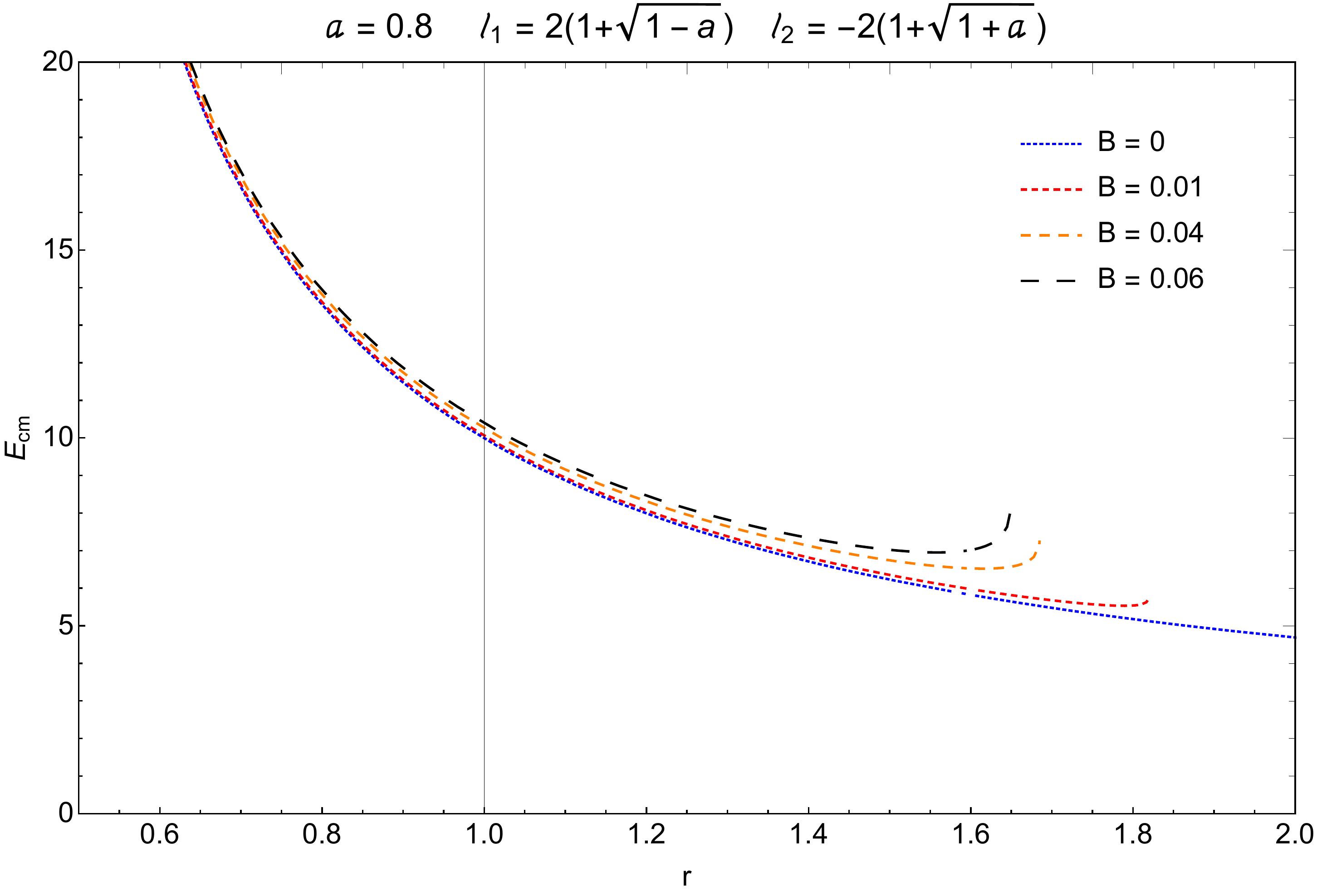} }}%
    \qquad
    \subfloat[]{{\includegraphics[width=8cm]{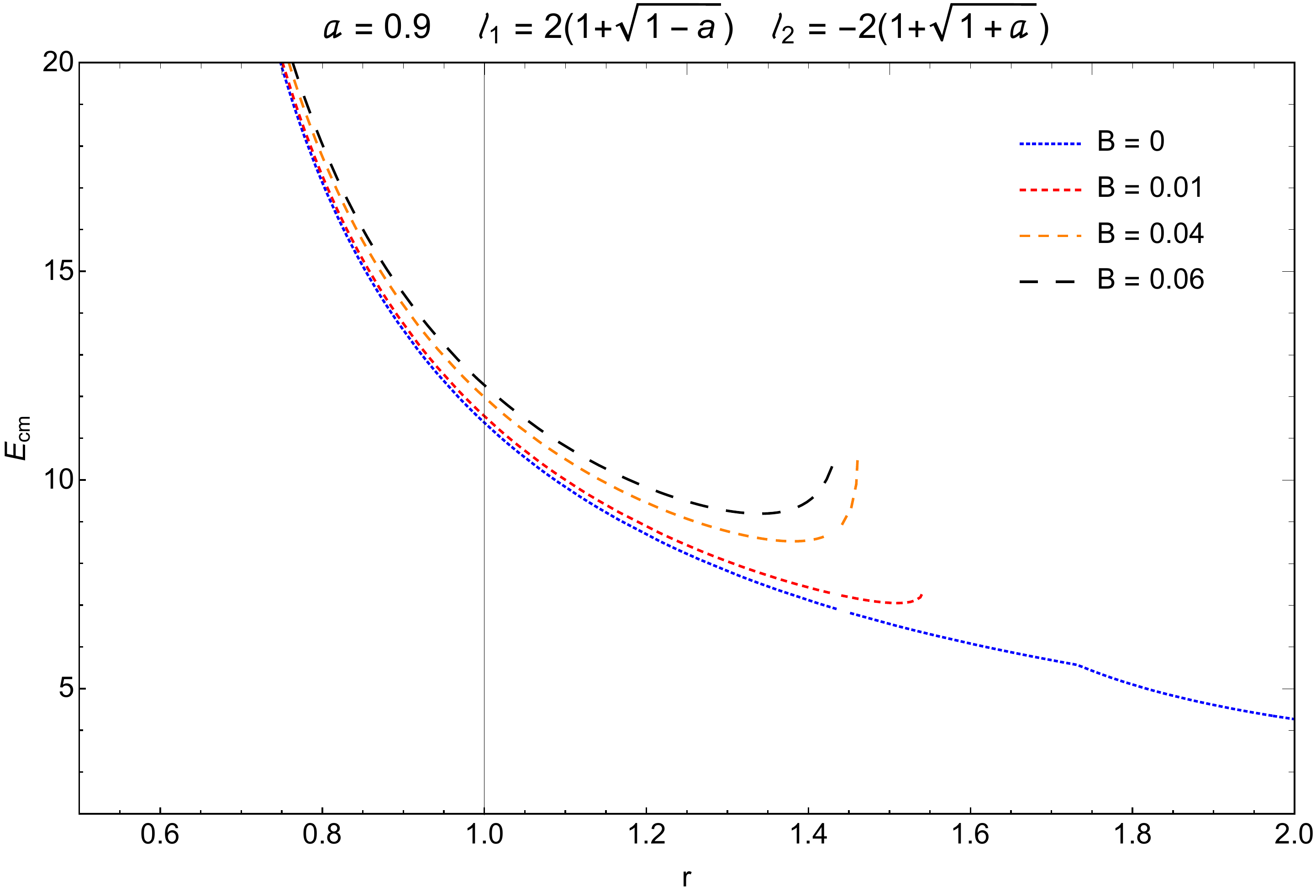} }}%
    \caption{Variation of $E_{c.m.}$ for (a) $a=0.8$ and (b) $a=0.9$, for four values of magnetic field $B$ in C.T/m units.}%
    \label{fig:fig9}%
\end{figure}

\begin{figure}[h]
    \centering
    \includegraphics[width=8cm]{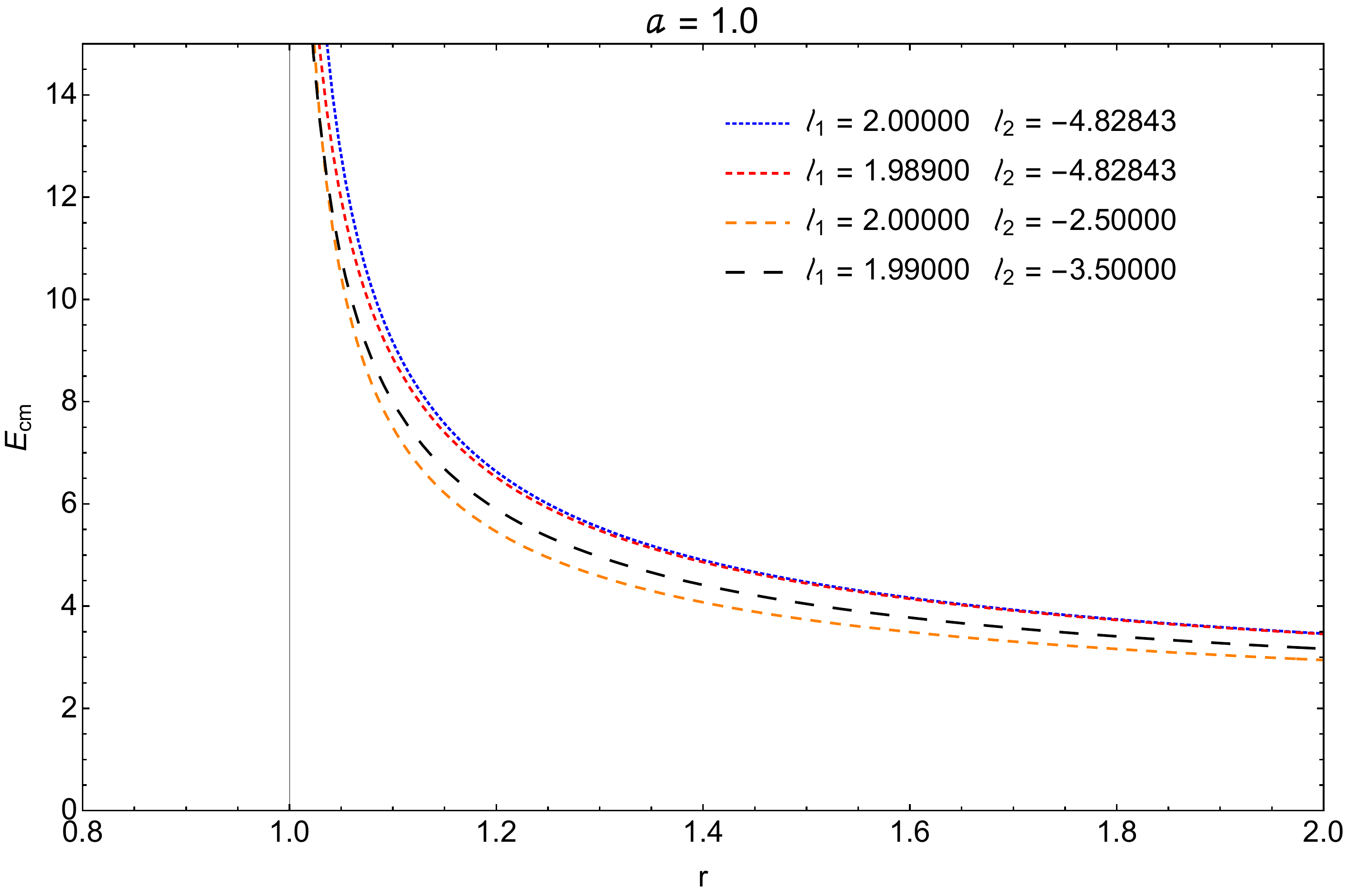}\\
\caption{Variation of $E_{c.m.}$ for $a=1.0$ (extremal black hole) for four different values of angular momentum for neutral particles. Before the horizon $r=1$, the colliding particles are accelerated to infinity, as awaited \cite{banados}. }
\label{fig:fig10}
\end{figure}

\begin{figure}[h]
    \centering
    \subfloat[]{{\includegraphics[width=8cm]{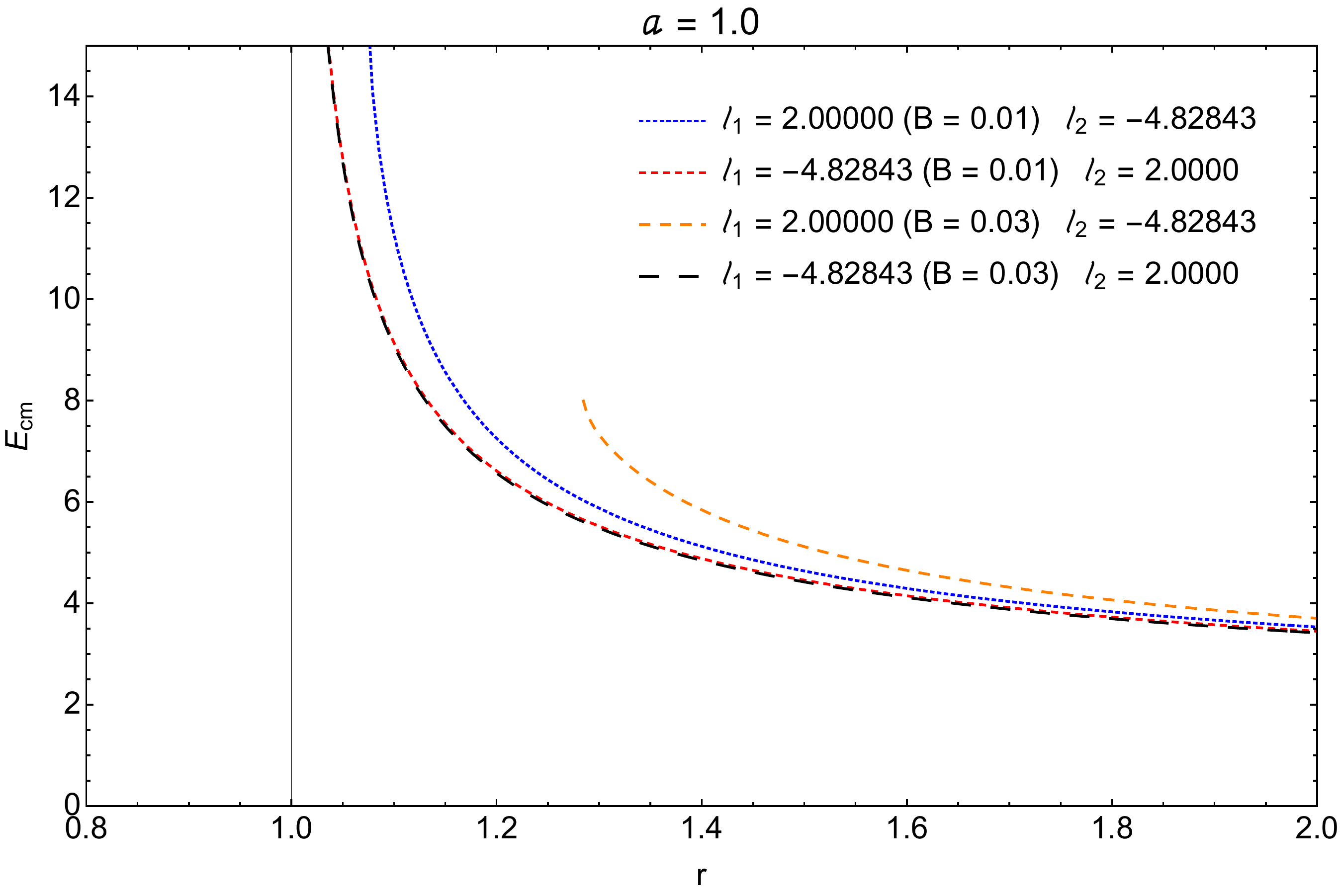} }}%
    \qquad
    \subfloat[]{{\includegraphics[width=8cm]{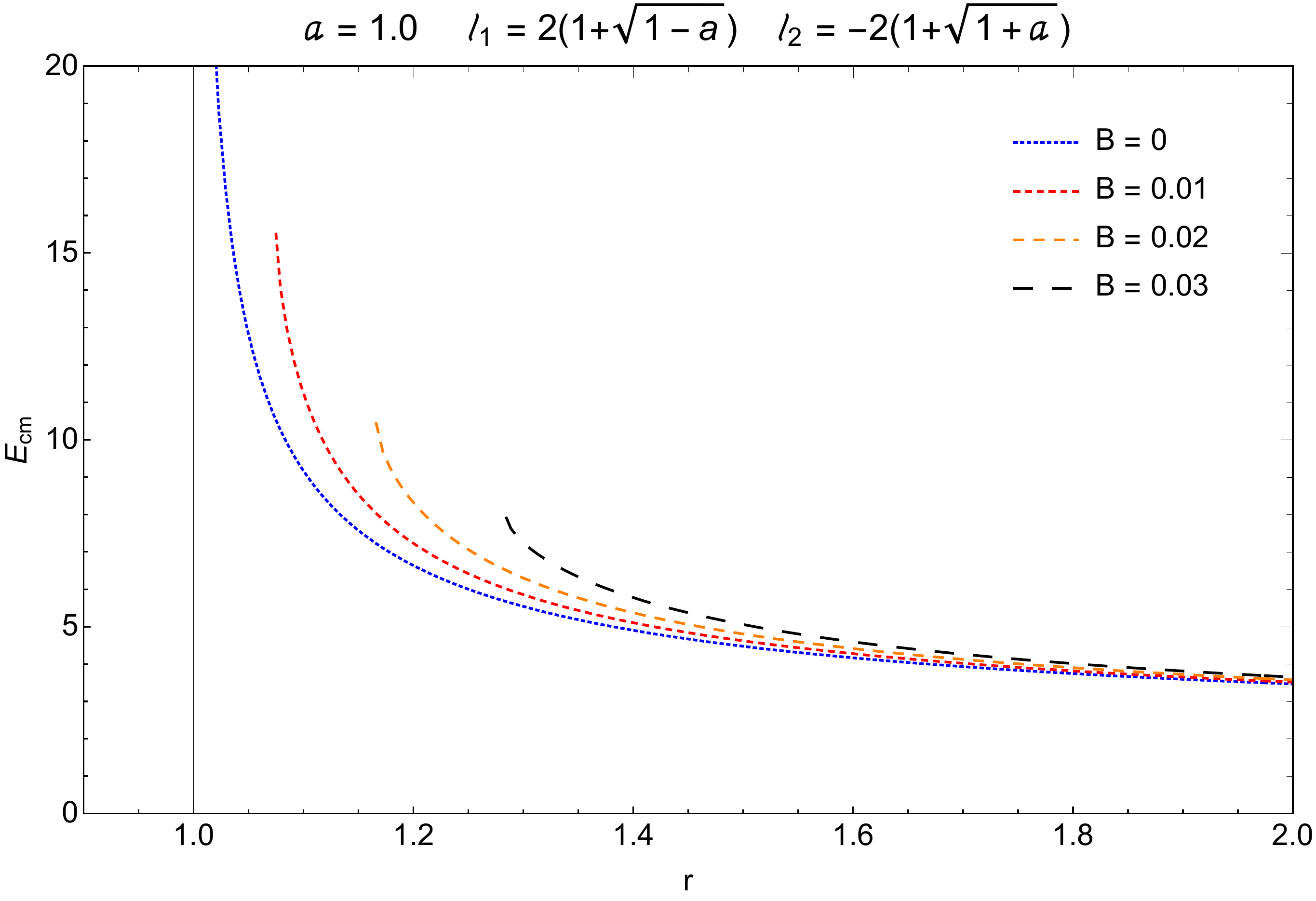} }}%
\caption{Variation of $E_{c.m.}$ in a extremal BH ($a=1$) for (a) neutral-charged particle collision for two values of magnetic field $B$ in C.T/m units and four different values of angular momentum and (b) for charged-charged particle collision. Before the horizon $r=1$, the particles are accelerated to infinity. }
\label{fig:fig11}
\end{figure}

\begin{figure}[h]
    \centering
    \includegraphics[width=8cm]{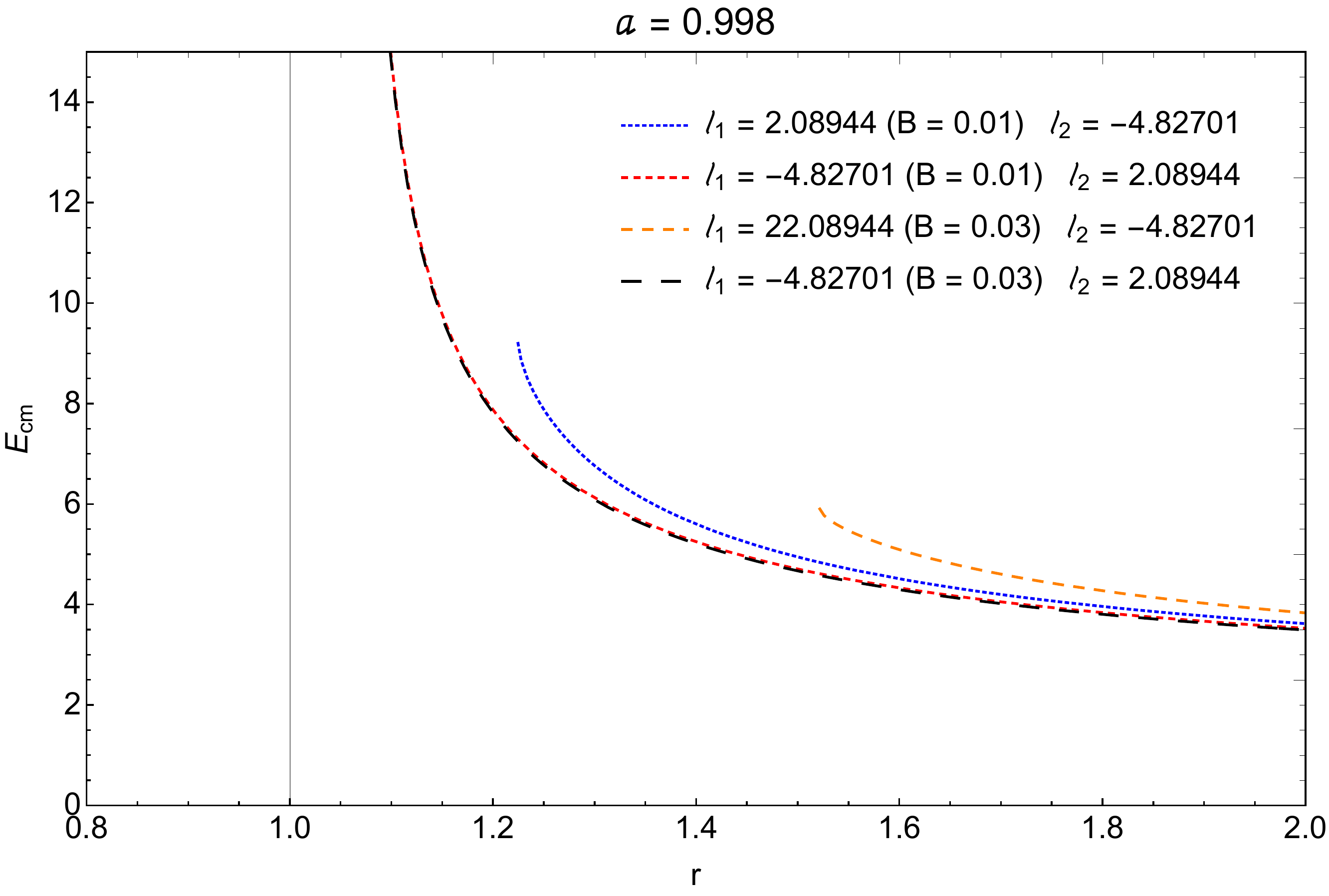}\\
\caption{Variation of $E_{c.m.}$ for $a=0.998$ (near-extremal black hole) for four different values of angular momentum for neutral-charged particles in the presence of $B=0.01$ and $B=0.03$ magnetic fields. Before the horizon, the colliding particles are accelerated to infinity. }
\label{fig:fig12}
\end{figure}

\begin{figure}[h]
    \centering
    \includegraphics[width=8cm]{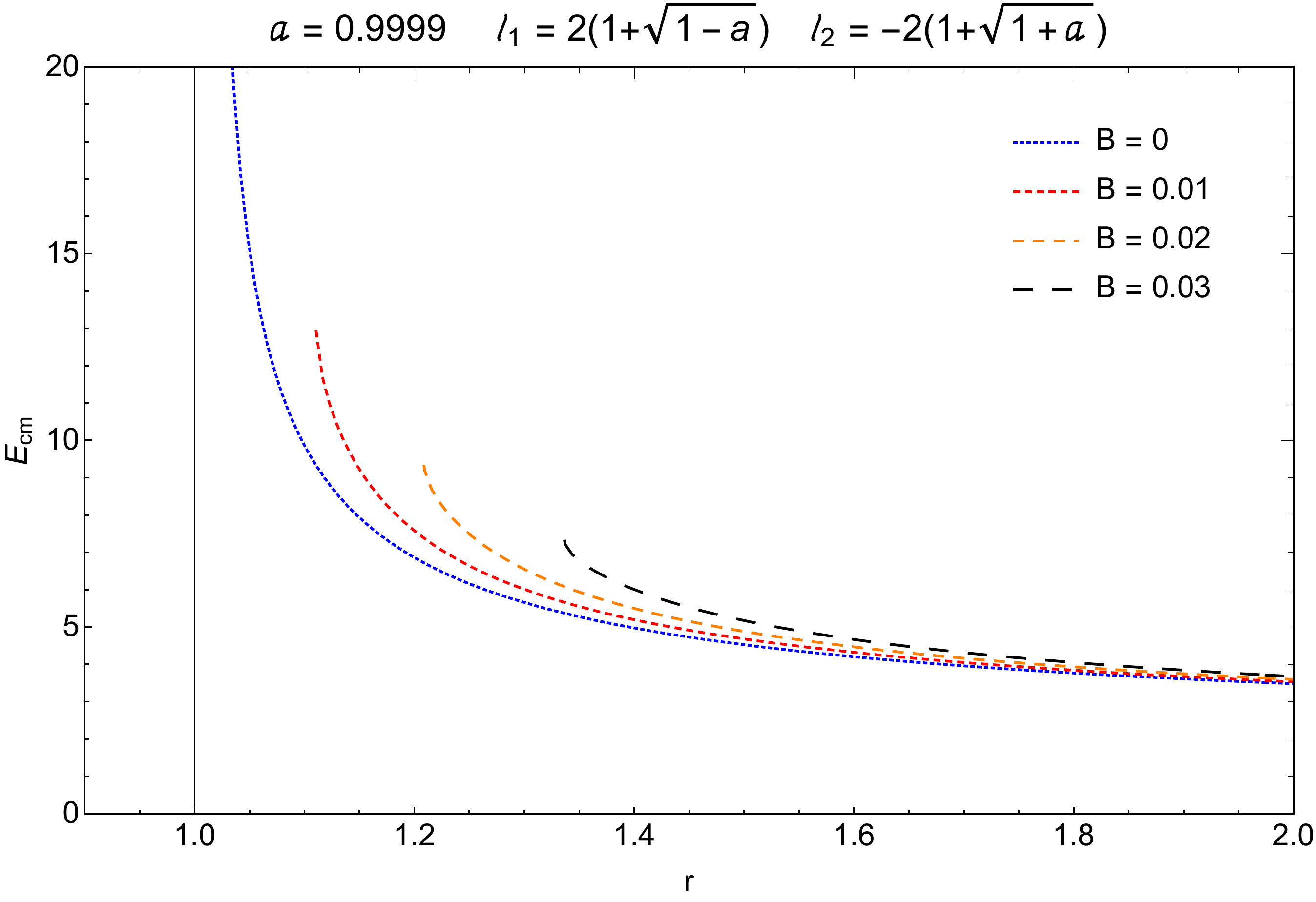}\\
\caption{Variation of $E_{c.m.}$ for a near-extremal black hole for four different values of magnetic fields for charged-charged particles. Before the horizon, the colliding particles are accelerated to infinity. }
\label{fig:fig13}
\end{figure}

\noi The plots of $E_{c.m.}$ energy are shown in Figs. \ref{fig:fig9}--\ref{fig:fig13} for some values of $a$ and $B$. Below we list the main results concerning $E_{c.m.}$:

\begin{itemize}
\item Regarding the case of particles (charged or neutral) in the vicinity of non-near-extreme black holes, $a < 0.998$, for any value of $B$, the particles cannot escape from the system, see Fig. \ref{fig:fig9}. 
\item Fig. \ref{fig:fig10}, for neutral-neutral colliding particles, shows that, for the $E_{c.m.}$ energy of particles that collide in the vicinity of extreme black holes ($a \rightarrow 1$), particles can in fact be accelerated escaping from the system (as awaited, see \cite{banados}).
\item Regarding the $E_{c.m.}$ energy of other type of particles that collide in the vicinity of extreme black holes ($a \rightarrow 1$), the results indicate that particles can in fact be accelerated escaping from the system, Fig. \ref{fig:fig11}a for neutral-charged and Fig. \ref{fig:fig11}b for charged-charged. 
\item Regarding the $E_{c.m.}$ energy of neutral-charged particles that collide in the vicinity of near-extremal black holes ($a \approx 0.998$), the results indicate that particles can in fact be accelerated escaping from the system (Fig. \ref{fig:fig12}).
\item Regarding the $E_{c.m.}$ energy of charged-charged particles that collide in the vicinity of near-extremal black holes, the results indicate that particles can in fact be accelerated escaping from the system (Fig. \ref{fig:fig13}).
\item The presence of magnetic fields of the order of magnitude of that observed in the center of AGNs ($\sim 10^2-10^4$ G, see Section \ref{section:3}) shows that neutral-charged or charged-charged particles could be accelerated at more distant locations in relation to the event horizon. 
\item Another point is that only a certain range of magnetic fields actually cause the acceleration of the particle in positions increasingly farther from the horizon (see Figs. \ref{fig:fig11} and \ref{fig:fig13}), confirming the issue commented about the behavior of $V_{eff}$ regarding magnetic fields. Otherwise, magnetic fields of great magnitude can cause the ISCO particle to lose its stability and, after collision, it will fall onto the horizon. 
\end{itemize}

\begin{figure}%
    \centering
    \subfloat[]{{\includegraphics[width=8cm]{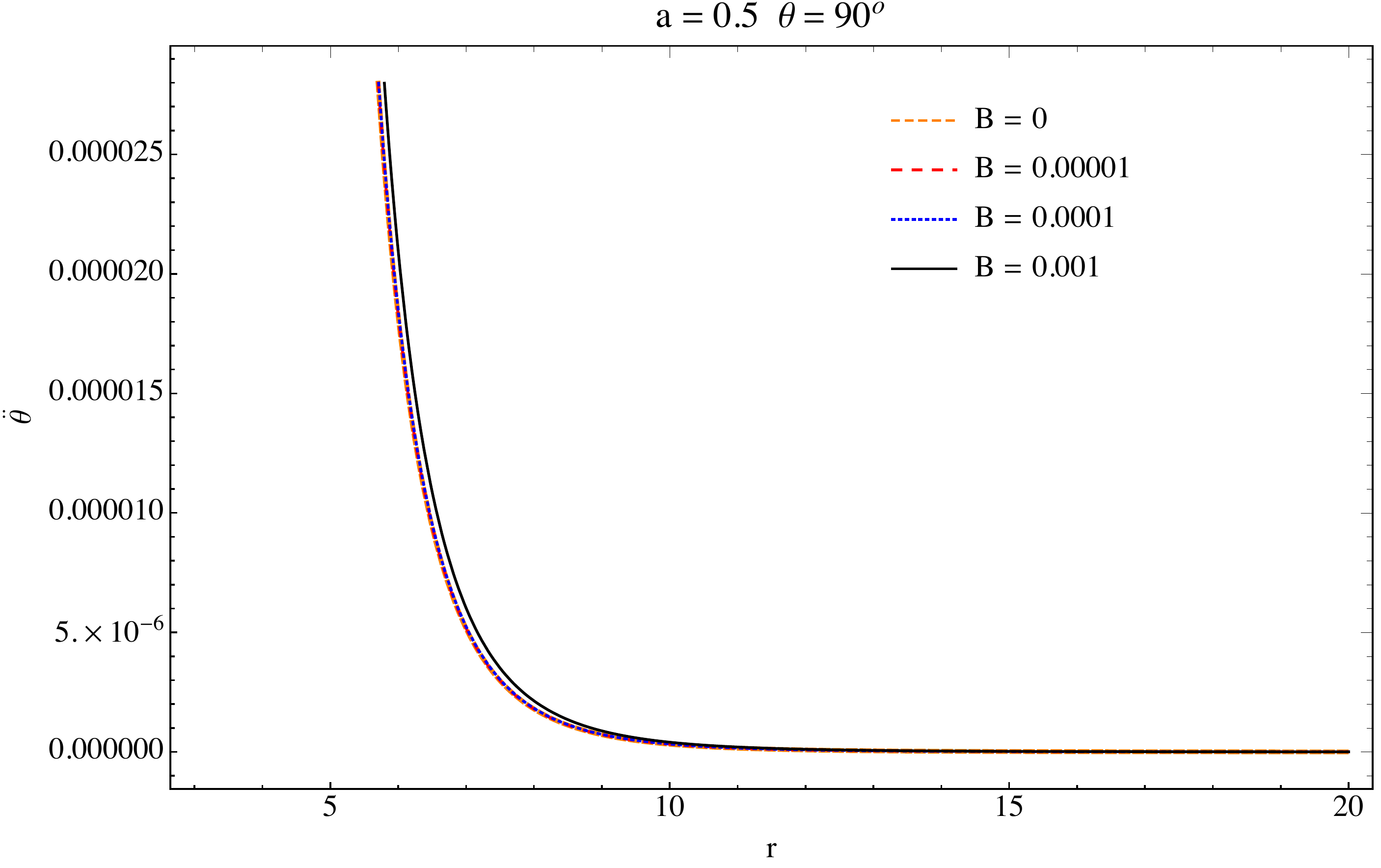} }}%
    \qquad
    \subfloat[]{{\includegraphics[width=8cm]{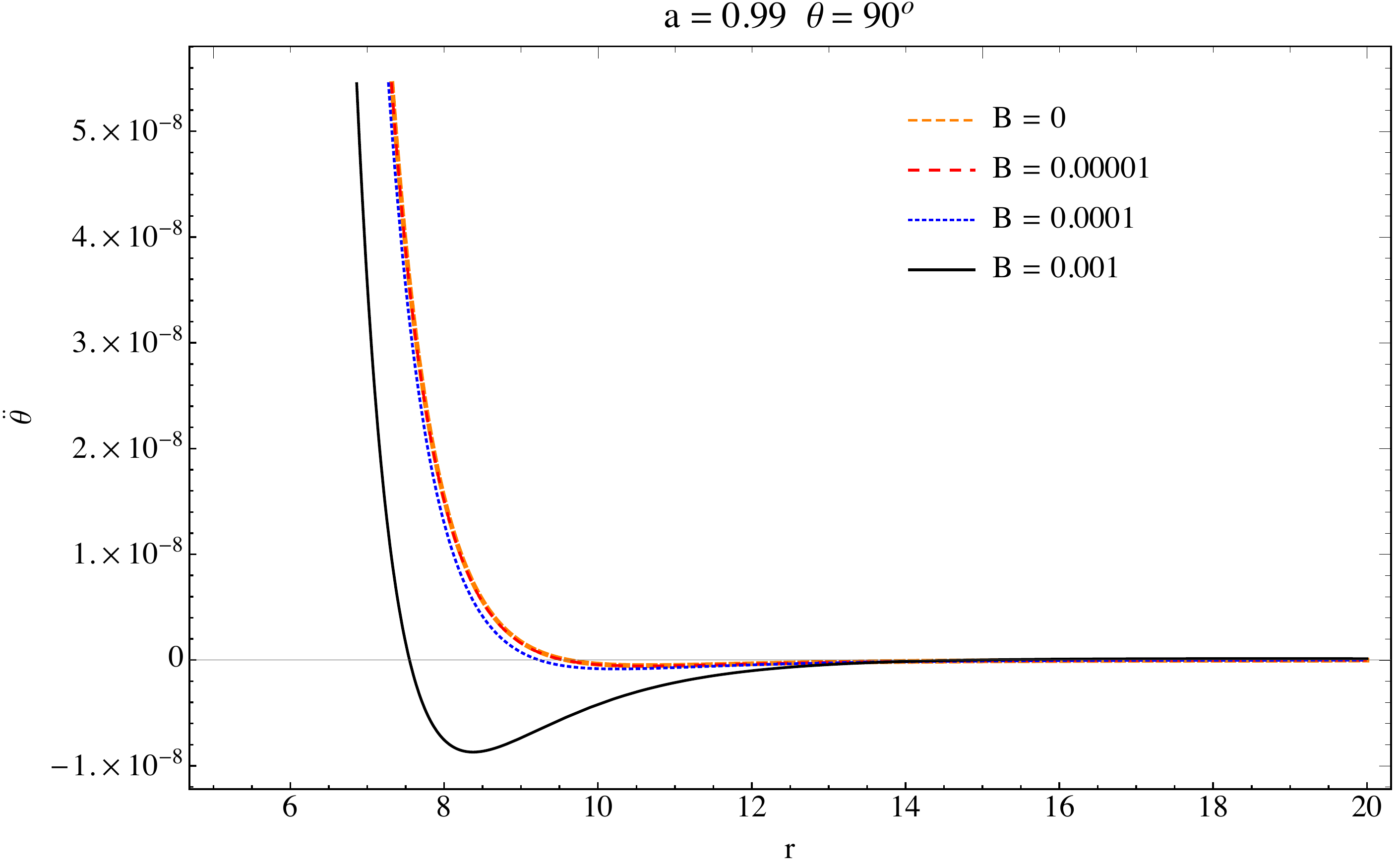} }}%
    \caption{Axial angular acceleration.}%
    \label{fig:fig14}%
\end{figure}

Thus, magnetic fields can contribute in the acceleration of particles, but above certain values of $B$, the particles lose stability and end up slowing down or being captured by the black hole itself. For example, $B$s above 0.01 indicate axial angular deceleration (Fig. \ref{fig:fig14}) which may in fact indicate the particles fall towards the horizon.

Another problem is the question about the real existence of extreme black holes. As discussed in \cite{thorne,gammie,volonteri,benson,bambi2,brenneman,shi1,gralla,shi2}, extreme black holes do not seem to really exist astrophysically, which makes dubious the existence of an astrophysical system composed only of extremal black holes and magnetic fields as real accelerators of charged particles. Restrictions on maximum $E_{c.m.}$ caused by gravitational radiation and back-reaction would solely permit infinite $E_{c.m.}$ only at infinite time and on the horizon of the black hole. Thus, the mechanism that accelerates UHECRs in AGNs is more likely to be linked to the jet mechanism and to classical mechanisms modeled for particle acceleration (Fermi, reconnection, etc). Although apparently black holes cannot reach the critical spin $a=1$, smaller values can be reached, as for example $a=0.998$ (near-extremal case) \cite{thorne}. In the present Section we have shown that, in this last case, for a range of values of $B$, neutral-charged or charged-charged colliding particles are accelerated to high values of $E_{c.m.}$ and particles escape quickly from the system. Magnetic field acts accelerating particles further and further away from the horizon and apparently mimics the role of the spin $a$. 

\section{Concluding remarks}\label{section:concluding}

In the present paper we have studied the motion of charged particles near magnetized rotating black holes and the related energetic processes in the vicinity of the Kerr BH horizon. It is investigated the collision of two charged particles falling freely from rest at infinity. Theoretically it is possible to extract all the rest energy of a mass by lowering it into a Schwarzschild BH, and even more energy applying a Penrose process lowering such mass into a Kerr BH \cite{blandford,jacobson}. This is the same to say that the rotation dynamics is, in principle, a more than sufficient source of energy for energizing powerful relativistic jets and, in consequence, high energy charged particles. In this respect, we have shown under what conditions particle can escape from the vicinity of the black hole to spatial infinity and the viability of magnetic extraction to drive the acceleration of charged particles.

The presented results follows similar conclusions in literature that points to the possibility that Kerr BHs could indeed act as particle accelerators. In this respect, we have discussed the possibility of degeneracy between the BH spin and the local magnetic field. Here it was showed that it appears that certain values of fine-tuned axial magnetic fields also could help in the process of acceleration. Otherwise, axial magnetic fields could cause a damping effect beyond certain $B$ values. In general, as showed in the present work, possibly only extremal rotating black holes with $a=1$ or near-extremal black holes with $a \approx 0.998$ (if magnetized) could behave as particle accelerators. Regarding this last specific case, important astronomical observations of specific spectral lines \cite{bambi2,brenneman,daly}, made by specialized telescopes as the Event Horizon Telescope (EHT) reveal that localized  emissivity  enhancement orbiting near a high-spin black hole are constraing and confirming the existence of near-extremal BH sources in the universe  (see, e.g., \cite{gralla,reynolds}).

\begin{acknowledgments}

The authors are very grateful to researchers of DEE-UFPR. The research of RCA is supported by CNPq grant numbers 307750/2017-5 and 401634/2018-3, and Serrapilheira Institute grant number Serra-1708-15022. The research of CHCA is supported by CNPq grant number 458896/2013-6. 

\end{acknowledgments}

\end{document}